# Prosocial Persuasion at Scale?

# Large Language Models Outperform Humans in Donation Appeals Across Levels of Personalization

John Pascal Caffier

Tilburg University

Olga Stavrova

Mannheim University & Tilburg University

Bennett Kleinberg

Tilburg University & University College London

## Author Note

Correspondence concerning this article should be addressed to John Pascal Caffier, Tilburg University. Email: J.P.Caffier@tilburguniversity.edu

**Abstract**

Large Language Models (LLMs) are increasingly regarded as having the potential to generate persuasive content at scale. While previous studies have focused on the risks associated with LLM-generated misinformation, the role of LLMs in enabling prosocial persuasion is still underexplored. We investigate whether donation appeals authored by LLMs are as effective as those written by humans across degrees of personalization. Two preregistered online experiments ($N_1$ = 658; $N_2$ = 642) manipulated Personalization (generic vs. personalized vs. falsely personalized) and Content source (human vs. LLM) and presented participants with donation appeals for charities. We assessed how participants distributed their bonus money across the charities, how they engaged with the donation appeals, and how persuasive they found them. In both experiments, LLM-generated content yielded statistically significantly higher donation amounts than human-authored content, although the absolute differences were small; it also resulted in higher engagement and was rated as more persuasive. There was a gain associated with personalization (Study 2) and a penalty for false personalization (Study 1). Our results suggest that LLMs may be a suitable technology for generating content that can encourage prosocial behavior.

## Introduction

The increasing public availability of Large Language Models (LLMs), such as ChatGPT (OpenAI, 2022), Gemini (Google, 2023), and Claude (Anthropic, 2023), has led to a rapidly developing field of research interested in evaluating their influence on individuals and society at large. This research has initially focused on LLMs' ability to successfully generate persuasive misinformation effectively (Goldstein et al., 2024; Mantello & Ho, 2024; Spitale et al., 2023). However, the potential of these new technologies for prosocial influence has not yet been sufficiently examined (e.g., Dörr et al., 2025). Further, it has been argued that the effectiveness of LLMs lies in their capacity to produce personalized content targeting individuals' sociodemographic profiles and/or personality traits (Teeny & Matz, 2024). Some research data are in line with those suggestions (Matz, Teeny, et al., 2024), whereas other work has raised doubt about the effectiveness of personalized approaches to persuasion (Hackenburg & Margetts, 2024a). Therefore, the potential of LLMs to promote prosocial behavior - and the extent to which personalization augments their persuasive power - remains to be determined. To address these research gaps, we explore whether donation appeals generated by LLMs differ in their effectiveness in yielding donations from human-written appeals. We further investigate how personalization influences the effectiveness of donation appeals by both humans and LLMs.

### Persuasion with Large Language Models

Persuasion - the process of influencing a person's opinion, attitude, or behavior through communication, for example, via exposure to text messages such as social media posts - has long been considered a uniquely human capacity. This has changed with the public roll-outs of generative LLMs (Argyle et al., 2025; Gao et al., 2023; Rogiers et al., 2024; Virtosu & Goian, 2023). Generative LLMs can create text content that readers often cannot distinguish from human writing (Barreto et al., 2023; Jakesch et al., 2023), a finding that extends to outputs of creative writing tasks, such as poetry production (Köbis & Mossink, 2021). Recent evidence further indicates that LLMs themselves can be susceptible to classic persuasion principles, responding in ways that resemble humans (Meincke et al., 2026).

The communication skills of LLMs have led scholars to express concerns about possible misuse (Kreps & Kriner, 2023), such as generating disinformation at scales unattainable by humans (Goldstein et al., 2024; Spitale et al., 2023; Huang & Wang, 2023). Indeed, in one study, LLM-generated fictional "news articles" containing disinformation on diverse topics (e.g., COVID-19, the Russia-Ukraine war, health, US elections, regional issues) were positively evaluated by humans in terms of text quality, narratives, and new arguments (Vykopal et al., 2023). Further studies demonstrated LLMs' success in direct opinion manipulation, for example, influencing the attitudes of citizens regarding policy issues and elections (like Brexit, gun bans, or carbon taxes; Bai et al., 2025).

Academics have recently raised the question of whether new technologies like LLMs can be capable of *prosocial* persuasion at scale as well (Dörr et al., 2025). Recent work suggested that personalized LLM dialogues reduced conspiracy convictions by approximately 20% among 2190 conspiracy believers, with effects lasting two months (Costello et al., 2024). Herein, we ask whether LLMs can achieve even more than that: Can they persuade humans to engage in costly prosocial behavior - charitable giving - as effectively as human persuaders?

The process of behavioral change is different from the process of attitude change: For example, engaging in a certain behavior usually has consequences, while articulating (or changing) a belief in an anonymous survey often does not. Prosocial behavior in particular is "costly" (Barclay & Barker, 2020; Verhallen & Pieters, 1984), as it implies donating money or time, going beyond the "cheap talk" of attitude expression. Prosocial behavior varies considerably in the costs it imposes on the actor: Whereas donating a small amount of money represents a relatively low-cost act, other forms of prosociality, such as long-term volunteering or making substantial personal sacrifices, may lead to far greater opportunity costs and life consequences (Schroeder, 2024). The present studies operationalize prosocial behavior at the lower end of the continuum, focusing on a low-threshold form of everyday prosocial action. For that, successfully persuading individuals to donate money mostly requires engaging with individuals' emotional and motivational dispositions, such as eliciting empathy, invoking moral identity, or triggering the feeling of guilt (Molho et al., 2025; Andreoni, 1990; Batson, 1991).

Are LLMs capable of this? This question is particularly non-trivial in the domain of prosocial behavior: as outlined above, persuading individuals to engage in that kind of behavior (e.g., donating) requires precisely those emotional and moral capacities that are traditionally considered the province of human communicators. While LLMs seem to use morally charged language more intensively than humans (Carrasco-Farre, 2024), they still score lower than humans in moral competence tests (Bajpai et al., 2025). At the same time, it is also observed that LLMs are getting better at correctly interpreting people's mental states (Zhu et al., 2024) and showing empathy (Kleinberg et al., 2024; Inzlicht et al., 2024). For example, even relationship advice offered by LLMs is sometimes evaluated as more human and empathetic than that provided by humans (Festor et al., 2026; Rubin et al., 2025; Yin et al., 2024). As LLMs are capable of generating texts that display empathy, they might also be capable of generating texts that bring out empathy from the readers, such as in donation appeals. Since empathetic concern is one of the central motivations for prosocial behavior (Bloom, 2017; Telle & Pfister, 2016; Eisenberg & Miller, 1987), LLMs could be well capable of creating persuasive and effective donation appeals. This is also supported by a recent preprint that suggests that personalized conversations with an LLM chatbot led to higher charitable donations compared with non-personalized conversations on an unrelated topic (White et al., 2024).

Although these early results are promising, it remains unclear whether LLMs' persuasiveness in the area of charitable giving is comparable to that of humans. The human – LLM comparison is important as humans are still the main force in the fundraising industry (Hahn et al., 2025; Koshy, 2025; Neumann, 2025). It has been demonstrated that LLM-generated texts in areas other than prosocial behavior can even be more effective or at least comparable to human persuaders (Schoenegger et al., 2025). Other studies show LLMs' lower perceived credibility (Ji & Zhang, 2026) or highlight the challenges in reliably simulating human behavior at the individual level (Petrov et al., 2024). Nevertheless, meta-analytic work shows that, on average, LLMs can be as persuasive as humans in fields like journalism and customer service (Huang & Wang, 2023). Moreover, policy-oriented experiments have demonstrated that arguments generated by LLMs can be even more effective in influencing public opinions than those crafted by humans (Bai et al., 2025). However, it remains unclear whether LLMs can motivate costly prosocial behaviors - such as charitable giving - as effectively as human writers. To date, no study has benchmarked LLMs' prosocial persuasion abilities directly against those of humans, even though LLM-human comparisons are standard practice in this line of research (e.g., Schoenegger et al., 2025). We acknowledge that, in practice, some charities' social media posts may be authored by niche-specialized communicators (such as fundraising agencies). Our studies do not benchmark LLMs against such niche-specialized humans (such as freelancers or agencies); instead, they compare LLM-generated appeals with appeals written by academically trained behavioral scientists (Study 1) and by financially incentivized lay writers from the target population (Study 2). This provides a theoretically and practically relevant first benchmark, as – depending on the organizational strategy, resources, and communication capacity (Nah & Saxton, 2013) – many prosocial appeals in social media will also in the future probably be written by volunteers of grassroots organizations, researchers, and administrative charity staff without specialized training in niche-specific fundraising copywriting. Moreover, this approach is consistent with other recent LLM-human benchmarking studies (e.g.,Bai et al., 2025; Schoenegger et al., 2025).

**Effect of personalization**

Personalization means systematically modifying the content, tone, word choice, and other communicative dimensions to match the characteristics of the intended recipient, e.g., demographic attributes, values, or identity-relevant traits. In contrast, generic content is not tailored to specific audience. We further distinguish accurate personalization from false personalization: falsely personalized appeals are tailored to different recipients' data and therefore create an audience-message mismatch.

Apart from invoking empathy to generate donations for charities, LLMs' donation appeals can be effective due to their ability to personalize persuasive messages for each recipient at virtually no cost. Several conceptual frameworks emphasize the possible advantages of audience-specific, personalized communication. For instance, according to the Elaboration Likelihood Model (ELM), an increased personal relevance can potentially motivate a deeper, more central processing of messages, which can increase persuasiveness (Lee et al., 2025; Chen & Lee, 2008; Cacioppo et al., 1986). Social Identity Theory also predicts that messages customized based on group memberships (e.g., political or religious ideology) can promote ingroup identification and trust, which then can also increase message effectiveness further (Turner et al., 1979). Donation appeals are perceived as more persuasive when aligned with individuals' values and customized to target their implicit and explicit preferences (Kesberg & Keller, 2021; Zarouali et al., 2022; Zettler & Strandsbjerg, 2025). In theory, by providing LLMs with individuals' personal data, such as values, preferences, or demographics, LLMs can create mass-scalable content that is personalized to each recipient (Teeny & Matz, 2024). Yet, empirical studies of the benefits of personalization in LLM persuasion provided mixed results (Teeny & Matz, 2024; Hackenburg & Margetts, 2024b). In a number of studies, LLM-generated personalized content based on demographic information was rated as more persuasive than generic (non-personalized) content (Matz, Beck, et al., 2024; Salvi et al., 2024; Teeny & Matz, 2024). Other studies, however, failed to link personalization to any benefits (Timm et al., 2025; Hackenburg & Margetts, 2024a). In sum, while many theoretical arguments speak in favor of personalization, the empirical evidence is mixed, and studies of the personalization effects in the context of prosocial persuasion are lacking.

**The present research**

We ran two online experiments to examine the persuasiveness of donation appeals written by Large Language Models (LLMs) versus those written by humans, across different levels of personalization. In both experiments, the study had a 2 (Content Source: human vs. LLM) × 3 (Personalization: personalized, generic, falsely personalized) within-subjects factorial design. In both studies, participants read six social media posts ostensibly written by different charities asking for donations. Participants rated each message for perceived persuasiveness, responded via Like, Dislike, or Neutral buttons, and could allocate a share of their bonus ($0.10) to the corresponding charities. In both studies, LLM messages were generated using *gpt-4.5-preview* (OpenAI, 2025). In Study 1, human messages were written by writers with an academic background in behavioral sciences, whereas in Study 2, human writers were laypeople from the U.S. who were asked to target their own demographic cluster for the personalized posts and were incentivized to write the posts that would generate the highest donation amounts.

## Study 1

The study was preregistered at the Open Science Framework (OSF; doi.org/10.17605/OSF.IO/XF95V) and was approved by the local ERB. Data, study materials, and an overview of deviations from the preregistration are available at osf.io/9rqax/overview?view_only=44a8e9faf046477cba0dbba2dc67d6f4.

### Method

#### *Participants*

A sample of $N$ = 750 participants was recruited via Prolific (Palan & Schitter, 2018). The target sample size was determined *a priori* via a simulation-based power analysis conducted in *R* with linear mixed-effects models fitted with the *lme4* R package (Bates et al., 2025).

The simulated dataset included six observations per participant (as in the actual study design), and the model included fixed effects for the two manipulated factors (content source and personalization) and their interaction, as well as random intercepts at the level of participants and charities. We used 1000 simulations and targeted small interaction effects ($\beta$ = .05). The results showed that a sample of 750 participants would be large enough to provide approximately 85% power to detect the specified above interaction effect at an *α* = .05 significance level.

The sample was quota-balanced to represent the US population in terms of sex, age, and political affiliation (targeted using Prolific target data options). Participants were only allowed to complete the study on a computer (and not on a smartphone or tablet) for a consistent device-related user experience. Participants provided their informed consent before participating and received financial compensation of $1.00, plus a bonus of up to $0.10 in accordance with their donation decisions. 725 participants fully completed the study, and out of them, 67 (9.2%) participants were excluded using the *careless* R package (Yentes & Wilhelm, 2023) after screening for an always-same response pattern on the persuasiveness items (identical persuasiveness ratings for each of the six posts; e.g., rating all three persuasiveness items of Post 1 as 3, all three for Post 2 as 4, etc.), producing a final analytic sample of 658 participants (Age: *M* = 45.05 years, *SD* = 15.77; Gender: 327 female, 331 male).

### ***Procedure***

We used a 3 (Personalization: personalized vs. falsely personalized vs. generic) × 2 (Content source: LLM-generated vs. Human-authored) within-subjects design. Each participant evaluated six social media posts, representing six different charities, one in each of the six experimental conditions. Charities were fully crossed with conditions: For example, the same charity was featured in the personalized LLM condition for one participant but in the generic human-written condition for another participant. To generate personalized texts, we collected participants' demographic data (age, gender (male/female), political ideology, and religiosity) in advance: Political ideology was measured on a six-point scale (1 = *Strongly Left/Liberal* to 6 = *Strongly Right/Conservative*). Religiosity was measured as well on a six-point scale (1 = *Deeply Non-Religious* to 6 = *Deeply Religious*). Based on these demographics, participants were allocated to one of 24 target group clusters (age: young [18–34], middle [35–54], older [55+]; gender: male, female; political views: left, right; religiosity: religious, not religious). For example, the cluster "OMRR" would refer to Older,

Male, Right-leaning, and Religious participants. Also, a falsely personalized profile code was created for each participant to assign them to falsely personalized posts: In the example cluster above, this would have been "YFLN" (Young, Female, Left-leaning, Non-religious). Middle-aged participants ("M") were randomly assigned the attribute "Old" or "Young" to assign them to falsely personalized posts.

Following the demographic collection and cluster assignment, each participant viewed six social media posts containing calls for donations. The posts were displayed within realistic social media wrappers that were modeled after the appearance of social media posts on Twitter/X for ecological validity (see Appendix 1). All posts were presented on a single page in a randomized order to avoid carryover effects. Importantly, participants were also not informed about the authorship or personalization of any post and were thus blinded to the experimental conditions. Participants were then asked to (a) allocate their promised bonus payment of $0.10 between themselves and/or the six charities, (b) engage with the post using “Like,” “Neutral,” or “Dislike” buttons, and (c) rate the persuasiveness using a short questionnaire. The order in which participants provided post engagement and persuasion ratings and made donation decisions was randomized across participants as well. Upon completion, the study ended with a debriefing that explained the study’s background and its goals. Participants were informed that the charities’ posts were fictional, the charities themselves are real, and their donation decisions would be executed by the research team.

***Stimuli Development***

The stimuli consisted of social media posts supposedly written by six real-world U.S.-based cancer research charities with politically and religiously neutral names. Human-written posts were written by graduate and postgraduate students with an academic training in behavioral sciences ($n$ = 16), and LLM-generated texts were generated using the latest model *gpt-4.5-preview* from OpenAI (OpenAI, 2025). For each personalization condition (personalized, generic, falsely personalized), both human authors and the LLM followed the same instructions while being restricted to a character limit of 280 characters (analogous to the Twitter/X character limit; Gligorić et al., 2018): *"Create a persuasive Twitter/X post encouraging donations to [charity name]. (In case of personalized condition: Tailor the tone, language, and message to resonate with this audience: [cluster profile, e.g., young, female, religious, left-wing].) Feel free to use emotionally engaging language and highlight relatable values or motivations, but avoid false claims. Include a clear call to action to donate now. Keep the text within 280 characters, including spaces."*

No LLM system prompt was used; the LLM was queried with the task instruction as the sole prompt message, and the temperature parameter and all other parameters were left at API defaults (see LLM Checklist at https://osf.io/rwaht/overview?view_only=e5175aa1b11d493faefd990451bbe4cf for details; Feuerriegel et al., 2026). In the generic condition, the audience-tailoring instruction parts (see above) were simply omitted; all other instructions remained identical across conditions.

Human writers were allowed to use the internet to gather information for their writing, but they were not allowed to use LLM tools. For each of the 24 target group clusters, two distinct versions of posts were created by human writers and LLM for each of the six charities, resulting in 600 posts (including 24 additional generic posts), and participants were randomly assigned to one version of the post within their respective experimental condition to make sure the results do not hinge on one specific writer or one specific LLM output. For internal evaluations, human writers were asked to provide feedback via Likert scales on the extent to which each personalization factor (e.g., gender, age) was utilized in their writing, for example, *"To what extent did you adapt the message to reflect the target gender (e.g., male, female)?"* for each post.

***Measures***

Prosocial behavior, our central dependent variable, was measured on the post level by participants' allocation of their $0.10 bonus across the six charities (representing the six experimental conditions). Post engagement was measured using a rating system where each social media post could be rated using the "Like"/"Neutral"/"Dislike" buttons (coded as 1, 0, and -1, respectively). Perceived persuasiveness was evaluated with three items on 7-

point Likert scales (1 = *Strongly Disagree* to 7 = *Strongly Agree*) adapted from Thomas et al. (2019): (1) “This social media post makes me want to donate”, (2) “The message in this social media post is convincing”, and (3) “This social media post provides a compelling reason to consider donating”; responses to these items were averaged to produce a single persuasiveness score per post (Cronbach’s $\alpha$ = .93).

***Analysis plan***

We performed separate linear mixed-effects model analyses for each of the outcome variables (donation amount, post engagement, persuasiveness), with fixed effects of Personalization Type (generic, personalized, falsely personalized), Content Source (human, LLM), and the personalization-by-content-source interaction. Random intercepts were added for participants and charities to account for the dependence of observations[1]. To verify that the donation results were not an artifact of the ipsative budget structure, we additionally conducted two robustness checks: a repeated-measures MANOVA on within-participant budget proportions (with paired follow-up contrasts) and a donation model that included each post's vertical position on the page (i.e., the order in which participants allocated the donations) as a fixed effect (to rule out a layout-driven position effect). These checks are reported alongside, and not in place of, the primary linear mixed-effects models.

When omnibus effects were significant, we conducted post-hoc pairwise contrasts with Tukey correction. All parameter estimates are reported with standardized coefficients and 95% confidence intervals. All analyses followed our preregistered plan using a significance level of $\alpha$ = .05. Intraclass Correlation Coefficients (ICCs) were calculated to determine the proportion of variance attributable to individual differences between participants and charities.

## Results

All 24 theoretically possible demographic clusters were observed in the final dataset, and their cluster distribution matched our target expectations for a representative US sample. 591 of 658 participants (89.8%) donated at least a part of their bonus. Spearman correlation analysis showed that the donation amount was positively correlated with post engagement ($\rho$ = .26, $p$ < .001) and persuasiveness ($\rho$ = .34, $p$ < .001). Means, standard deviations, correlations, and confidence intervals can be found in Table 1.

[1] The random intercept variance for participants in the donation model was estimated as zero, due to minimal between-subject variability in donation amounts at the per-charity level.

Table 1

*Means, standard deviations, and correlations with confidence intervals (Study 1)*

| **Variable** | ***M*** | ***SD*** | **1** | **2** | **3** | **4** | **5** | **6** |
|---|---|---|---|---|---|---|---|---|
| 1. Age | 45.05 | 15.76 | | | | | | |
| 2. Political Ideology | 3.47 | 1.50 | .13**<br>[.10, .16] | | | | | |
| 3. Religiosity | 3.59 | 1.56 | .15**<br>[.11, .17] | .39**<br>[.37, .42] | | | | |
| 4. Familiarity with charity | .24 | .62 | .01<br>[-.02, .05] | .09**<br>[.05, .12] | .12**<br>[.09, .15] | | | |
| 5. Donation Amount | .01 | .02 | -.03<br>[-.06, .01] | .08**<br>[.05, .12] | .20**<br>[.17, .23] | .15**<br>[.12, .18] | | |
| 6. Engagement | .59 | .64 | -.03<br>[-.06, .01] | .07**<br>[.03, .09] | .13**<br>[.10, .16] | .07**<br>[.03, .10] | .26**<br>[.23, .29] | |
| 7. Persuasiveness | 4.96 | 1.58 | -.03<br>[-.06, .00] | .10**<br>[.07, .14] | .29**<br>[.27, .32] | .16**<br>[.12, .18] | .34**<br>[.31, .37] | .43**<br>[.41, .46] |

*Note.* * $p$ < .05, ** $p$ < .01. Familiarity with charity corresponds to a question at the end of the study, whether the charity was known ("Did you know these charities before participating in the study?" with answer options "Yes, I knew this charity before", "No, I was not aware of this charity before", and "Not sure")

Figure 1

*Donation Amount, Engagement, and Persuasiveness (Study 1)*

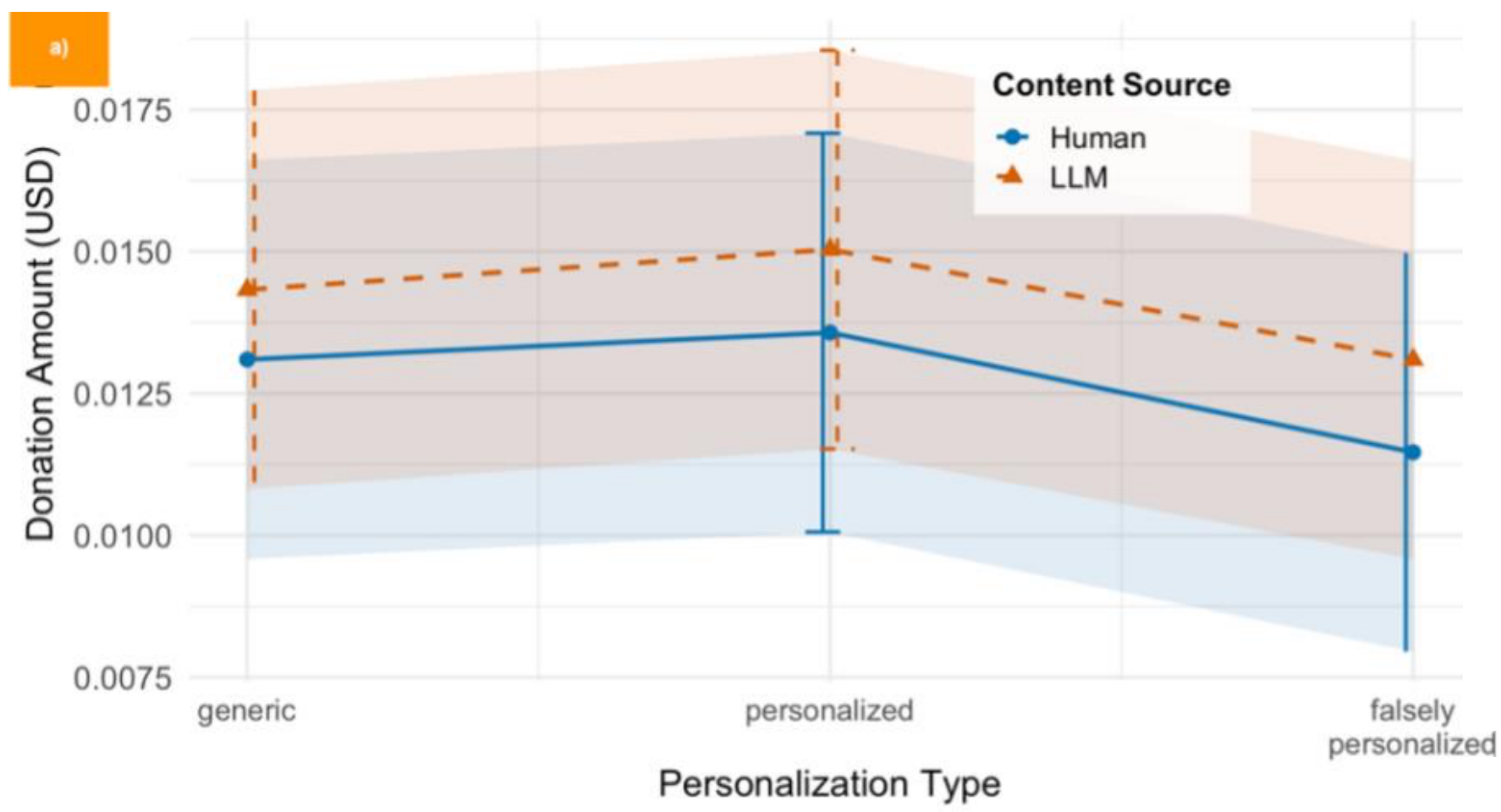

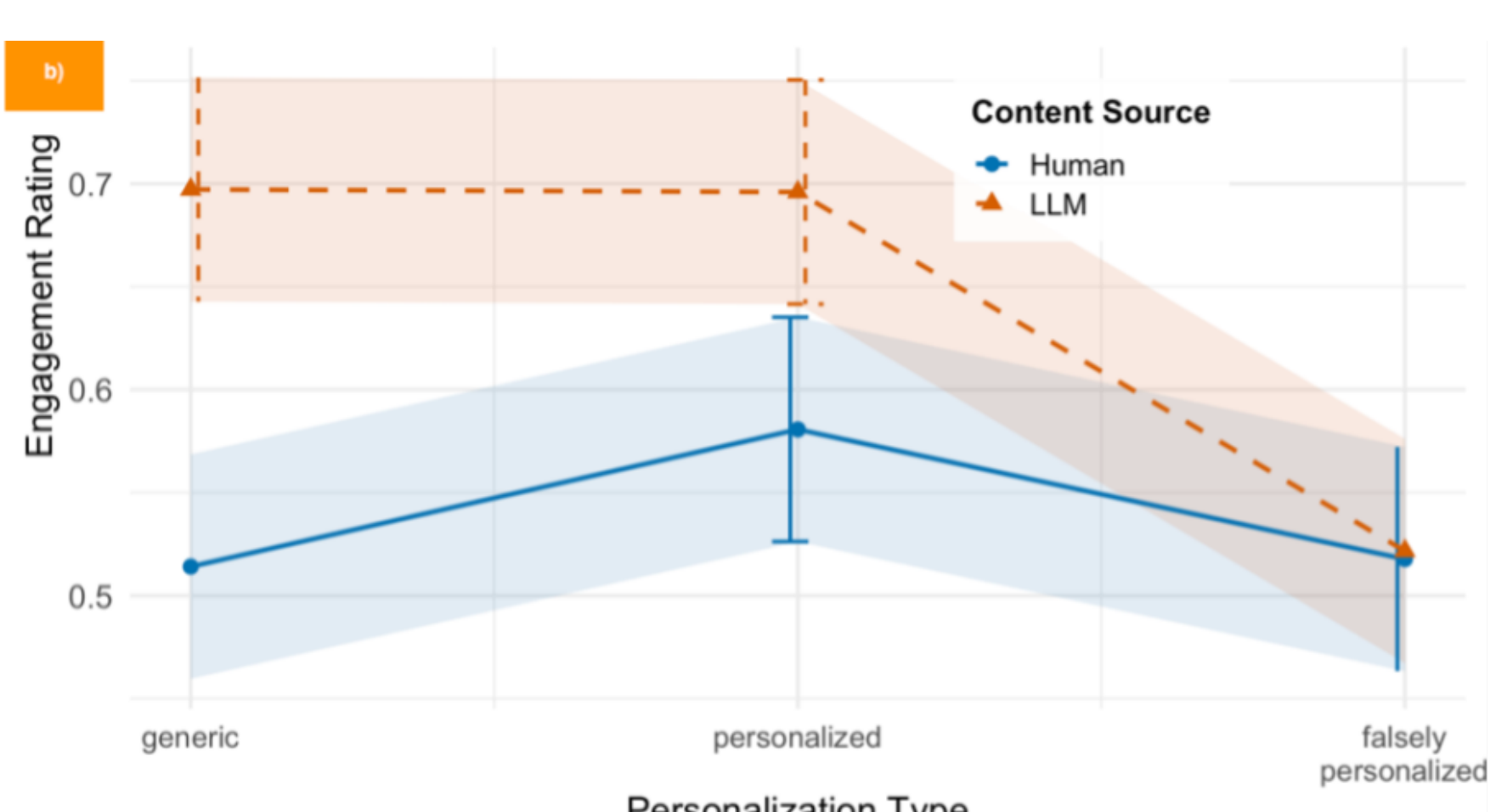

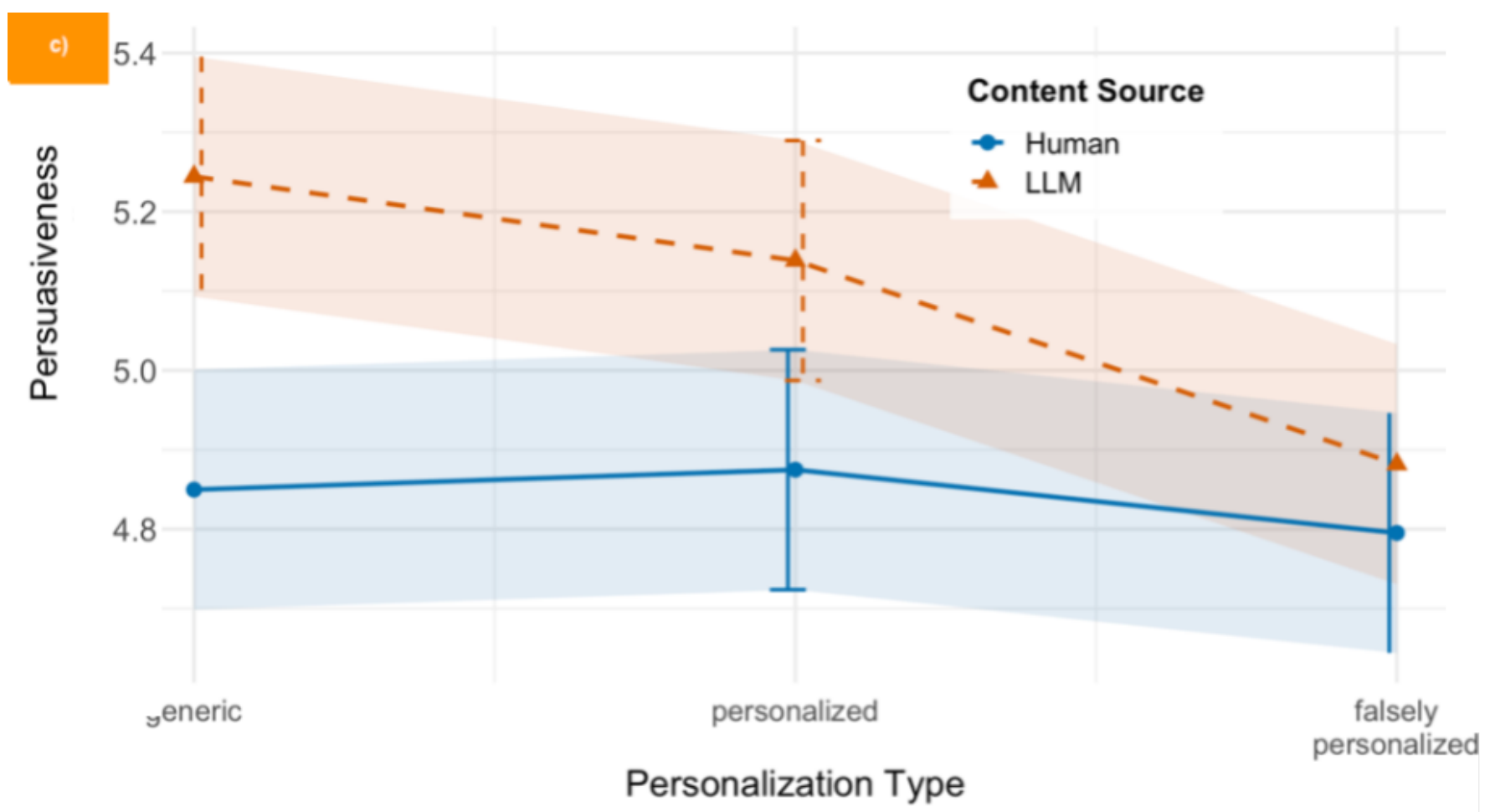

*Note.* Panel (a) displays mean donation amounts (USD; relative allocation differences within the fixed $0.10 bonus budget constraint). Panel (b) shows mean post engagement (-1 = *Dislike*, 0 = *Neutral*, 1 = *Like*). Panel (c) shows mean perceived persuasiveness on a 7-point Likert scale (1 = *Strongly Disagree* to 7 = *Strongly Agree*). Error bars and shaded bands represent 95% confidence intervals.

***Donation Behavior***

The multilevel model regressing donation amount on Personalization (generic, personalized, falsely personalized), Content Source (human, LLM) and their interaction suggested a significant main effect of content source, $F(1, 3939.2) = 6.05$, $p = .014$, $\eta^2_p = .002$, and of personalization type, $F(2, 3937.4) = 4.20$, $p = .015$, $\eta^2_p = .002$. The two factors did not interact significantly, $F(2, 3937.5) = .04$, $p = .961$. On average, participants donated a significantly higher percentage of their bonus budget to LLM-generated posts (see Appendices; *M* = 14.2%, *SD* = 19.1) compared to human-authored posts (*M* = 12.7%, *SD* = 18.4; d = .11, 95% CI [.02, .20], $t(3939.2) = 2.46$, $p = .014$). Regarding the effect of personalization, follow-up comparisons using Tukey adjustment showed that falsely personalized messages yielded significantly lower donations than personalized messages ($d = .09$, $p = .013$). Neither personalized nor falsely personalized messages significantly differed from generic messages ($d = .02$, 95% CI [−.04, .08], $p = .688$ and $d = .05$, 95% CI [.01, .11], $p = .113$, respectively).

Because the fixed bonus budget renders the six donations within a participant mutually dependent (partially ipsative), we probed the robustness of the donation results in two ways. First, we aggregated the data to the participant level and re-expressed donations as budget shares: for each participant who donated a nonzero amount ($n = 591$), we computed the share of their own donated total allocated to each of the six design cells (Content Source × Personalization Type), yielding six shares per participant that sum to one. These shares remain fully ipsative; the analysis, however, accommodates this dependence. Specifically, we conducted a repeated-measures MANOVA that tests within-participant contrasts between the shares and estimates its error covariance from the data, including the negative correlations among the outcomes created by the fixed budget (i.e., allocating more to one charity necessarily leaves less for others). The analysis thereby tests whether participants allocated *relatively* more of their donated budget to some conditions than to others. The pattern of results matched the primary analysis: a significant main effect of content source, *F*(1, 590) = 5.71, *p* = .017, a marginal main effect of personalization, *F*(2, 589) = 2.68, *p* = .069, and a nonsignificant interaction, *F*(2, 589) = 0.33, *p* = .717. Paired follow-up contrasts on participants' marginal shares showed that LLM appeals received a larger mean share of participants' donated budgets than human appeals (52.7% vs. 47.3%; mean paired difference = 5.3 percentage points, 95% CI [0.9, 9.7], *t*(590) = 2.39, *p* = .017, $d_z$= 0.10), along with personalized appeals a larger share than falsely personalized appeals (mean paired difference = 3.8 percentage points, 95% CI [0.3, 7.3], *t*(590) = 2.16, *p* = .031, $d_z$ = 0.09). Second, we tested a specific behavioral pathway through which the budget constraint could bias the condition estimates: budget depletion across the display. Because all six posts appeared simultaneously on one page in randomized order, participants scanning from top to bottom might assign more to earlier posts simply because more budget remained. We therefore added each post's vertical position as a fixed effect to the primary mixed-effects model (full analytic sample, *n* = 658). Donations indeed declined slightly from top to bottom (*b* = -0.0005, *p* = .002), consistent with mild depletion, but because position was randomized with respect to condition, the condition estimates were virtually unchanged (content source: *b* = 0.00144 in the primary model vs. 0.00140 with position included; falsely personalized: *b* = −0.00143 vs. -0.00142). This check does not remove the budget dependence itself (that is the purpose of the share-based analysis above) but rules out that the dependence manifested as an order-driven artifact confounded with condition.

***Post Engagement***

For post engagement, the multilevel model showed significant main effects for Personalization Type, *F*(2,3284.8) = 20.06, *p* < .001, $\eta^2_p$ = .012, and Content Source (*F*(1,3283.7) = 40.77, *p* < .001, $\eta^2_p$ = .012). LLM-generated content (*M* = .64, *SD* = .61) led to significantly higher post engagement than human-authored (*M* = .54, *SD* = .66) posts (*d* = .20, *t*(3283.7) = 6.39, *p* < .001). There was also a significant interaction between the two factors, *F*(2, 3283.3) = 11.01, *p* < .001, $\eta^2_p$ = .007. Tukey-adjusted post-hoc comparisons showed that for human-authored posts, personalized appeals received significantly higher engagement than generic appeals (*d* = .13, *p* = .039; 95% CI [.01, .26]). There were no significant differences between generic and falsely personalized posts, nor between personalized and falsely personalized posts. For LLM-generated posts, both generic (*d* = .32, *p* < .001, 95% CI [.21, .43]) and personalized appeals (*d* = .36, *p* < .001, 95% CI [.22, .48]) received significantly higher post engagement than falsely personalized appeals (both *p* < .001). There was no difference between generic and personalized LLM appeals.

### *Persuasiveness*

For perceived persuasiveness, the analysis showed significant main effects for Personalization type ($F$(2,3283.3) = 14.11, $p$ < .001, $\eta^2_p$ = .009) and Content source ($F$(1,3282.7) = 53.39, $p$ < .001, $\eta^2_p$ = .016). LLM-generated posts (*M* = 5.09, *SD* = 1.07) were rated significantly more persuasive than human-written posts (*M* = 4.84, *SD* = 1.56, *d* = .23, $t$(3282.7) = 7.31, *p* < .001). There was a main effect of personalization, *F*(2, 3283.3) = 14.11, *p* < .001. Post-hoc comparisons (Tukey-adjusted) showed that falsely personalized appeals were rated as less persuasive than both generic (*p* < .001) and personalized appeals (*p* = .0002), whereas generic and personalized messages did not differ (*p* = .596). A significant interaction between Personalization Type and Content Source was also observed (*F*(2, 3281.2) = 6.91, *p* = .001). Tukey-adjusted post-hoc comparisons showed that for human-authored posts, persuasiveness ratings did not differ significantly between generic, personalized, and falsely personalized appeals. In contrast, for LLM-generated posts, falsely personalized appeals were rated as significantly less persuasive than both generic appeals (*d* = .34, *p* < .001, 95% CI [.24, .46]) and personalized appeals (*d* = .24, *p* < .001, 95% CI [.11, .37]).

## Discussion

Study 1 explored whether LLMs may be capable of *prosocial* persuasion at scale. The main result of our first study was that donation appeals generated by an LLM were more effective than those written by humans in generating donations, eliciting post engagement, and persuasiveness. In addition, we also tested whether LLM-generated content can be even more convincing in promoting prosocial behavior when personalized to participants' socio-demographic profiles. Personalization had no significant advantage over generic messages in increasing donation amounts, while false personalization consistently reduced donation amounts, especially for LLM-generated content.

The observed superiority of LLM-generated over human-authored texts in eliciting prosocial behavior may be sample-specific and needs validation in other populations and contexts. Thus, in Study 2, we implemented several modifications in order to test the generalizability of the findings of our first study. Importantly, we tested whether LLMs' advantage emerges even under conditions designed specifically to increase the performance of human writers: In Study 2, human writers were recruited from the same population as study participants (U.S.), they were financially compensated, they wrote donation appeals targeting only *their own* (rather than randomly selected one) demographic profile, and were financially incentivized to draft the posts that would generate most donations.

## Study 2

Study 2 was preregistered at the Open Science Framework (OSF; doi.org/10.17605/OSF.IO/3wn9h). Data and study materials are available on osf.io/vu3xs/overview?view_only=e0d527cbc3c34524a066741fea0151e7.

### Method

#### *Participants*

*N* = 682 participants were recruited via Prolific using the same criteria as in Study 1. Following the same exclusion criteria as in Study 1 and our pre-registration, we removed 40 participants (5.9%) who were detected as careless responders. The final analytic sample then consisted of 642 participants (Age: *M* = 46.90 years, *SD* = 16.33 years; Gender: 340 Female, 302 Male). The sample was quota-balanced on political affiliation, sex, and age to represent the U.S. population distribution. The participants received $1.00 for participation, plus a bonus of up to $0.10 based on their donation amount.

***Procedure***

The experimental design and the procedure were the same as in Study 1.

***Stimuli Development***

As human writers, we recruited 164 U.S.-based users of Prolific (with the same U.S.-representative targeting as later participants), who wrote a total of 327 posts. For the personalized posts, they were asked to write donation appeals specifically for their own demographic cluster and also received a baseline compensation of $1.00 and competed for an additional $10 bonus awarded to the authors of the top 10 most effective posts. For the LLM-generated posts, the ones of Study 1 generated via *gpt-4.5-preview* were reused. The instruction for human writers was *"In this study, we will ask you to write several short donation requests on behalf of different charities. We will evaluate all requests with respect to the donation amount they will generate in our next study. The authors of the top 10 requests will each receive an additional $10 bonus payment. Try to be as persuasive as possible. [For personalized condition: Think about what will convince you to donate and tailor your message to effectively appeal to SOMEONE LIKE YOU.] Feel free to use emotionally engaging language and highlight relatable values or motivations, but avoid false claims. Include a clear call to action to donate now. Keep the text within 280 characters, including spaces. You can use search engines if needed. Generative AI use (e.g., ChatGPT, Gemini etc.) is not allowed."*

**Results**

As in Study 1, the sample collected in Study 2 comprised all demographic clusters. In total, 496 of 642 participants (77.3%) donated at least a portion of their bonus. On average, participants donated 71% of their bonus ($M$ = .071, $SD$ = .042). The average donation per charity post was $M$ = .012 ($SD$ = .022). Comparable to Study 1, donation amounts correlated with engagement ($r_s$ = .31, $p$ < .001) and perceived persuasiveness ($r_s$ = .38, $p$ < .001), see Table 2 for more details.

Table 2

*Means, standard deviations, and correlations with confidence intervals (Study 2)*

| **Variable** | ***M*** | ***SD*** | **1** | **2** | **3** | **4** | **5** | **6** |
|---|---|---|---|---|---|---|---|---|
| 1. Age | 46.90 | 16.33 | | | | | | |
| 2. Political Ideology | 3.30 | 1.49 | .12**<br>[.09, .15] | | | | | |
| 3. Religiosity | 3.14 | 1.72 | .12**<br>[.09, .15] | .45**<br>[.43, .48] | | | | |
| 4. Familiarity with charity | .17 | .59 | .00<br>[-.03, .03] | -.02<br>[-.05, .01] | .04*<br>[.01, .07] | | | |
| 5. Donation Amount | .01 | .02 | -.03<br>[-.06, .01] | .06**<br>[.03, .04] | .17**<br>[.14, .20] | .15**<br>[.12, .18] | | |
| 6. Engagement | .49 | .66 | -.04*<br>[-.06, .01] | .01<br>[-.02, .04] | .12**<br>[.09, .15] | .08**<br>[.04, .10] | .31**<br>[.28, .34] | |
| 7. Persuasiveness | 4.63 | 1.67 | -.03<br>[-.06, .00] | .02<br>[-.01, .05] | .19**<br>[.17, .23] | .13**<br>[.08, .15] | .38**<br>[.35, .41] | .49**<br>[.47, .51] |

*Note*. * $p < .05$, ** $p < .01$. Familiarity with charity was assessed as in Study 1.

Figure 2

*Donation Amount, Engagement, and Persuasiveness (Study 2)*

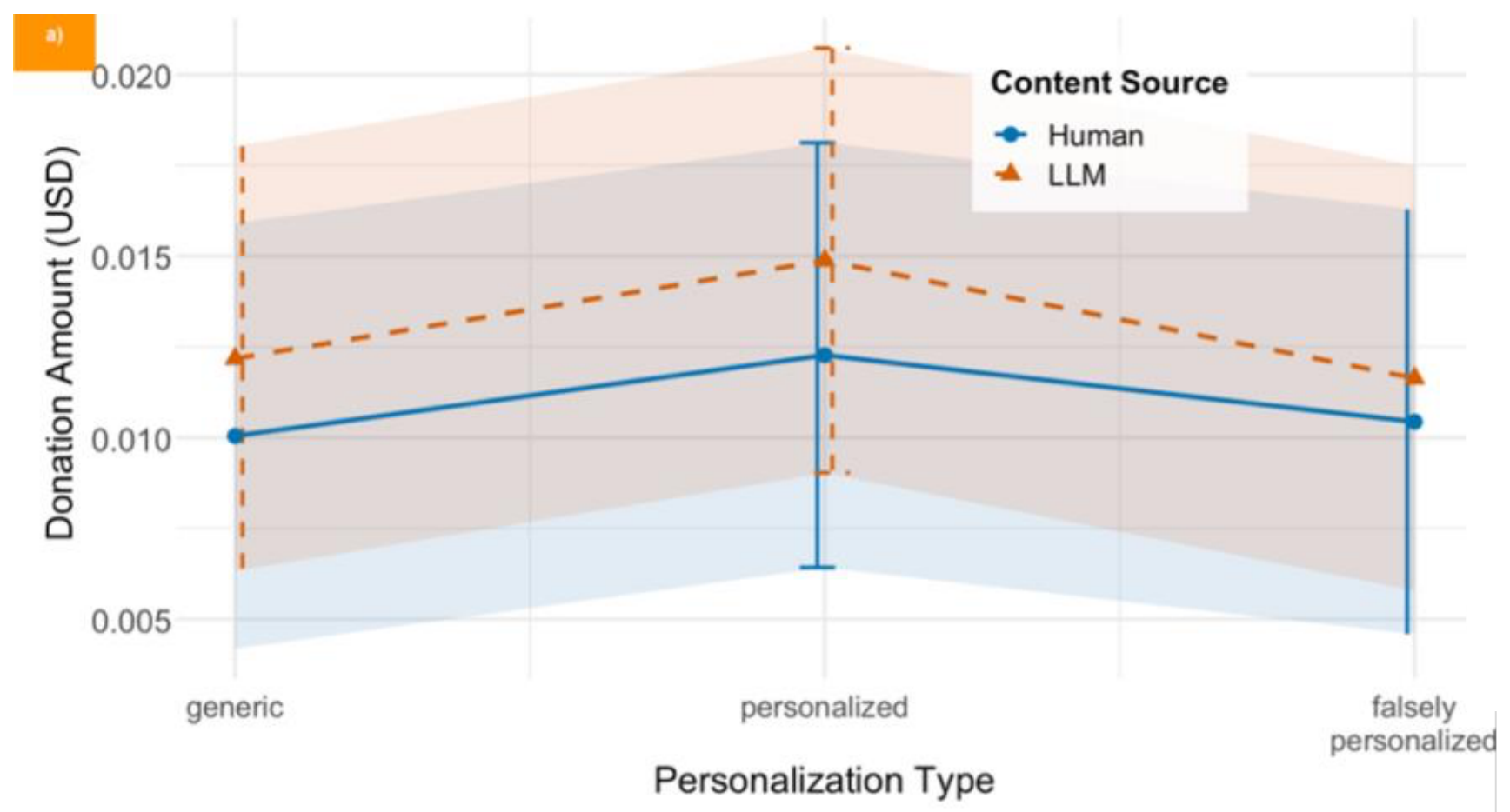

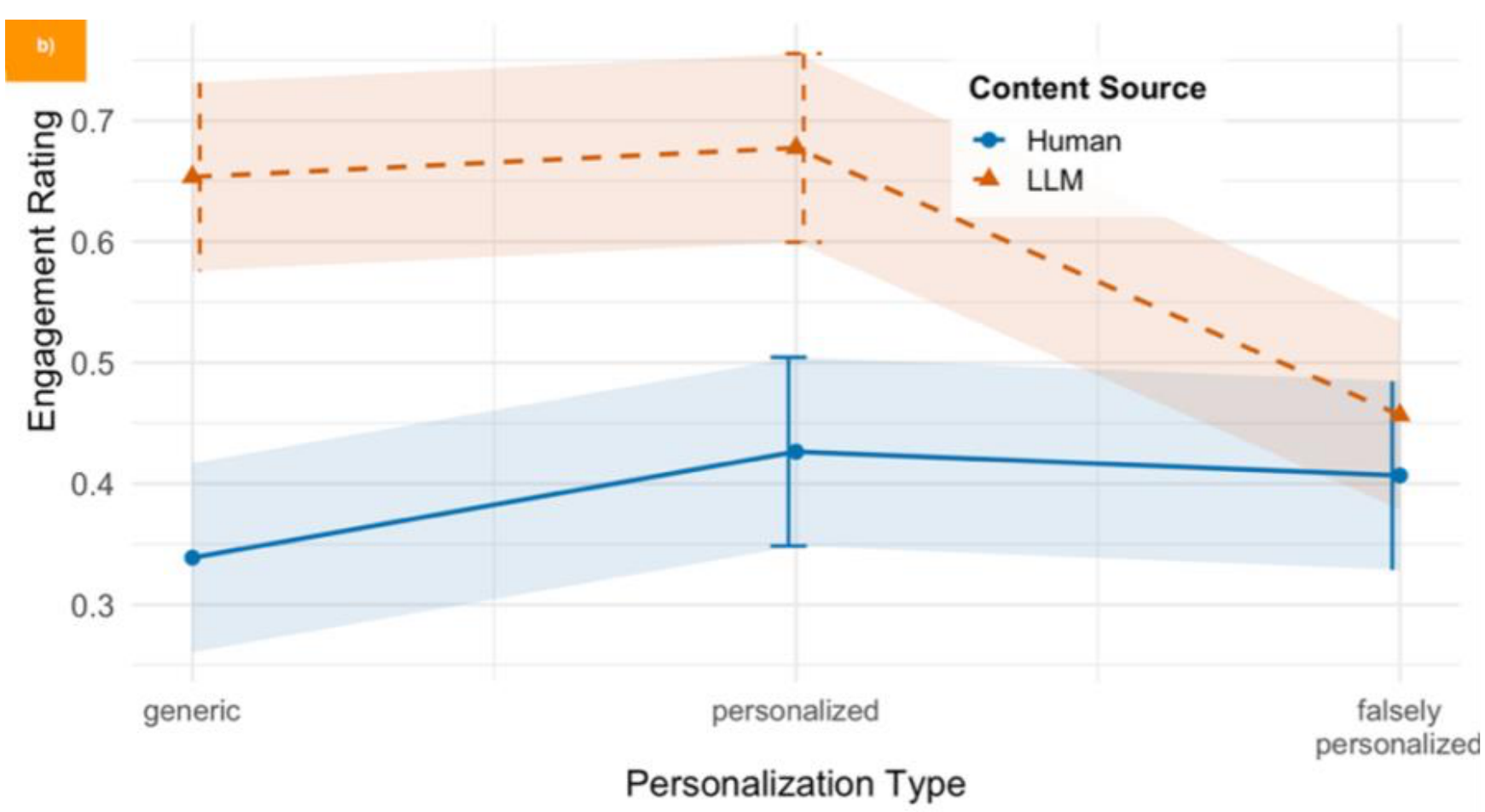

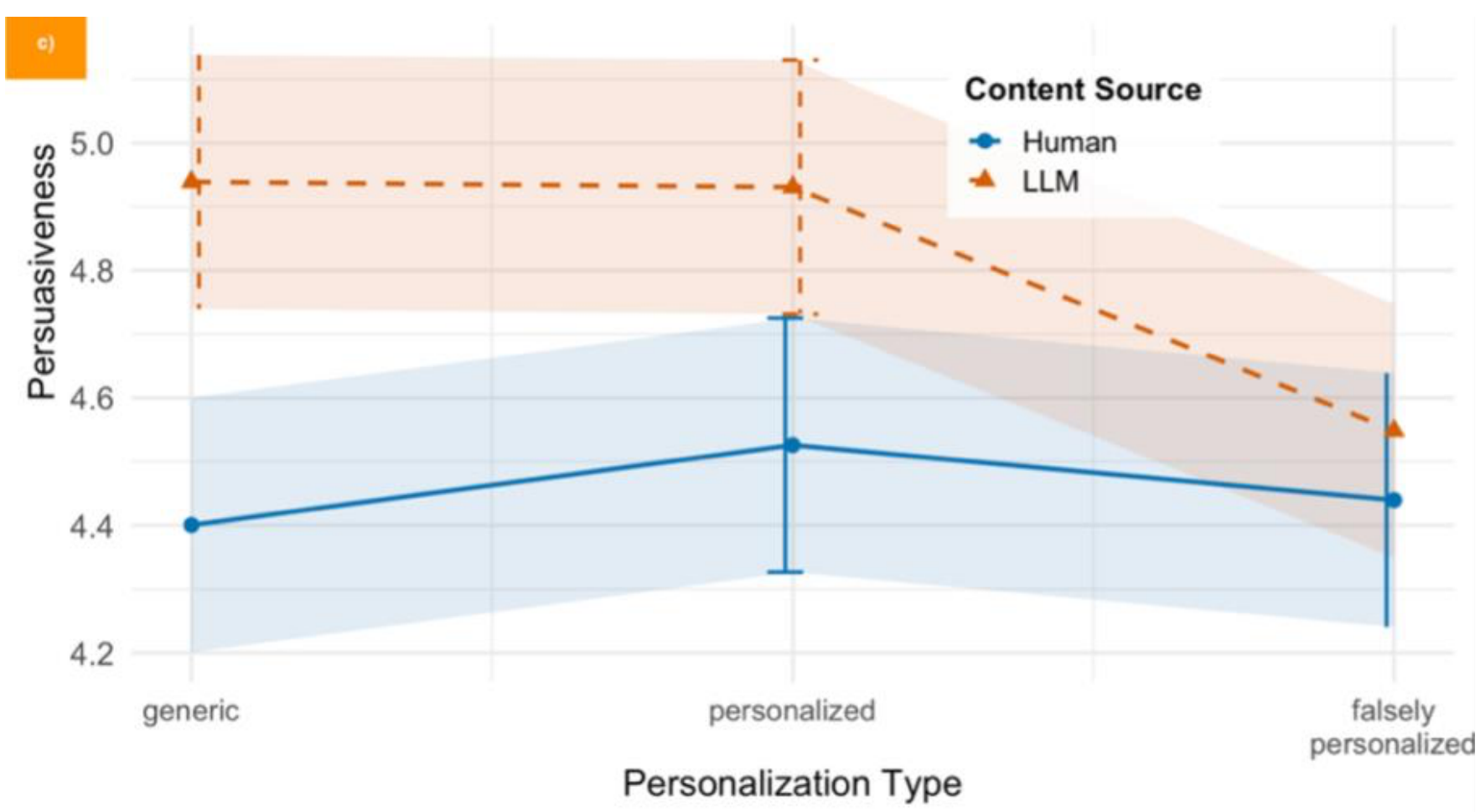

*Note.* Panel (a) displays mean donation amounts (USD; relative allocation differences within the fixed $0.10 bonus budget constraint). Panel (b) shows mean post engagement (-1 = *Dislike*, 0 = *Neutral*, 1 = *Like*). Panel (c) shows mean perceived persuasiveness on a 7-point Likert scale (1 = *Strongly Disagree* to 7 = *Strongly Agree*). Error bars and shaded bands represent 95% confidence intervals.

***Donation Behavior***

As in Study 1, a linear mixed-effects model was used to predict the donation amount. The analysis showed a significant main effect of content source, $F(1, 3844.1) = 8.24$, $p = .004$, $\eta^2_p = .002$, with LLM-generated appeals delivering higher donation amounts than human-authored appeals: LLM-generated posts received $M = 13.3\%$ ($SD = 23.6$; $M = .013$, 95% CI [.011, .015]), whereas human-written posts received $M = 10.5\%$ ($SD = 20.8$; $M = .01$, 95% CI [.009, .013]) of participants' bonus (for the full distribution of donation amounts, see Appendices). Also, as in Study 1, there was a significant main effect of personalization type, $F(2, 3841.5) = 5.90$, $p = .003$. But unlike in Study 1, personalized appeals received the highest proportion of the endowment ($M = 13.7\%$), followed by generic ($M = 11.1\%$) and falsely personalized appeals ($M = 11.0\%$). Tukey-adjusted post-hoc comparisons showed that personalized appeals produced significantly higher donations than generic appeals ($d = .12$, 95% CI [.03, .21], $p = .010$) and falsely personalized appeals ($d = .13$, 95% CI [.04, .22], p = .007). An interaction between the content source and the personalization type was not significant ($p = .69$).

We repeated both robustness checks on the donors in Study 2 ($n = 496$): The participant-level budget shares (six shares per participant, one per design cell, summing to one; donors only, $n = 496$) again confirmed the primary analysis: significant main effects of content source, $F(1, 495) = 11.63$, $p < .001$, and personalization type, $F(2, 494) = 6.10$, $p = .002$, and a nonsignificant interaction, $F(2, 494) = 0.91$, $p = .405$. In paired follow-up contrasts on participants' marginal shares, LLM appeals received a larger mean share of participants' donated budgets than human appeals (55.3% vs. 44.7%; mean paired difference = 10.6 percentage points, 95% CI [4.5, 16.7], $t(495) = 3.41$, $p < .001$, $d_z = 0.15$), and personalized appeals a larger share than both generic appeals (mean paired difference = 7.2 percentage points, 95% CI [2.3, 12.2], $t(495) = 2.86$, $p = .004$, $d_z = 0.13$) along with falsely personalized appeals (mean paired difference = 8.3 percentage points, 95% CI [3.3, 13.2], $t(495) = 3.29$, $p = .001$, $d_z = 0.15$). Second, adding each post's vertical position to the primary mixed-effects model (full analytic sample, *n* = 642) showed no position effect either ($b$ = -0.00002, $p$ = .939).

### *Post Engagement*

For post engagement, the analysis showed significant main effects for personalization type, $F(2, 3201.7) = 17.25$, $p < .001$, and content source, $F(1, 3203.7) = 149.14$, $p < .001$. Descriptively, generic appeals ($M$ = .49, 95% CI [.46, .53]) together with personalized appeals ($M$ = .55, 95% CI [.52, .58]) led to greater engagement than falsely personalized ones ($M$ = .43, 95% CI [.40, .46]). Tukey-adjusted post-hoc comparisons showed that personalized appeals were more engaging than falsely personalized appeals ($d$ = .12, $p < .001$) and generic appeals ($d$ = .06, $p$ = .007), and that generic appeals also led to higher engagement than falsely personalized appeals ($d$ = .06, $p$ = .002). As in Study 1, the interaction was also significant, $F(2, 3202.3) = 22.74$, $p < .001$. For human-authored posts, personalized appeals ($M$ = .43, 95% CI [.35, .50]) resulted in significantly higher engagement than generic appeals ($M$ = .34, 95% CI [.26, .42], $p$ = .008). Falsely personalized posts ($M$ = .41, 95% CI [.33, .49]) did not differ significantly in engagement compared to generic posts ($p$ = .051). There was no significant difference between personalized and falsely personalized human-authored posts ($p$ = .78). For LLM-generated posts, both generic ($M$ = .65, 95% CI [.58, .73]) and personalized ($M$ = .68, 95% CI [.60, .76]) appeals significantly did better compared to falsely personalized appeals in terms of engagement ($M$ =.46, 95% CI [.38, .54], $p < .001$ for both contrasts). The difference between generic and personalized LLM posts was not significant ($p$ = .69) for LLM-generated posts.

### *Persuasiveness*

We detected significant main effects of personalization, $F(2, 3201.4) = 13.67$, $p < .001$, $\eta^2_p = .007$, and content source, $F(1, 3205.1) = 83.19$, $p < .001$, $\eta^2_p = .021$, and also a significant interaction, $F(2, 3202.0) = 11.10$, $p < .001$, $\eta^2_p = .006$. LLM-generated posts ($M$ = 4.81; 95% CI [4.62, 4.99]) were rated as more persuasive than human-authored posts ($M$ = 4.46; 95% CI [4.27, 4.64]). Post-hoc contrasts showed that for human content, personalization did not significantly affect the persuasiveness ratings (generic: $M$ = 4.40 [4.20, 4.60]; personalized: $M$ = 4.53 [4.33, 4.73]; falsely personalized: $M$ = 4.44 [4.24, 4.64]; all pairwise comparisons $p > .14$). In contrast, for LLM content, falsely personalized appeals ($M$ = 4.55, 95% CI [4.35, 4.75]) were rated significantly less persuasive than generic ($M$ = 4.94, 95% CI [4.74, 5.14], $p < .001$) and personalized appeals ($M$ = 4.93, 95% CI [4.73, 5.13], $p < .001$).

### Discussion

Overall, the results of Study 2 are in line with Study 1: LLM-generated donation appeals performed better than human-authored ones across all outcome measures (donation behavior, post engagement, and persuasiveness). Unlike in Study 1, personalized posts generated higher donation amounts than generic and falsely personalized ones, across both sources. The results show that a well-implemented personalization is able to increase persuasion and promote prosocial behaviors like donating. Falsely personalized posts again performed worse than the generic ones, particularly for LLM content, suggesting that mismatched personalization backfires. The LLM advantage in donation behavior was descriptively larger in Study 2 than in Study 1 with lower average donations to human-written appeals, despite financial incentives and self-targeting for human writers in Study 2; one possible explanation is that academically trained writers in Study 1 had stronger general writing or argumentation skills than lay writers in Study 2, even though the latter benefited from greater personal familiarity with their target audience and stronger extrinsic motivation to perform well.

## Exploratory Analysis of Text Characteristics of Human- vs. LLM-Authored Appeals (Study 1 and 2)

To better understand potential linguistic differences between human- and LLM-authored donation appeals, we compared the human- and LLM-authored appeals on a range of linguistic characteristics. The unit of analysis was each individual appeal as used in both studies. Readability was assessed using the Flesch Reading Ease score (lower values indicate more difficult texts) and the Flesch–Kincaid Grade Level (estimated U.S. school grade required for comprehension). Lexical complexity was quantified using mean word length, the proportion of long words (>6 letters), type–token ratio (the proportion of unique words relative to all words), and a number of Linguistic Inquiry and Word Count software (LIWC-22) categories. LIWC represents a closed dictionary-based text analysis tool that estimates the prevalence of psychologically meaningful word categories (Tausczik & Pennebaker, 2010). Here, we focused on emotional, and moral language, as well as indices of writing style (Analytic, Clout, Authenticity, Dictionary coverage). Group differences between human- and LLM-authored appeals were evaluated using Welch's t-tests with Cohen's d as the standardized effect size. p-values were Holm-corrected across the 21 indices within each study.

Human and LLM appeals differed in these linguistic features across both independent human samples (academically trained writers in Study 1; incentivized lay writers in Study 2). The LLM produced on average ~36-word long appeals (*SD* = 2.7); Human-written appeals varied in length across the two writer samples: In Study 1, appeals written by academically trained writers were longer than those generated by the LLM (*M* = 38.7 vs. 36.0 words; *d* = -0.44). In contrast, in Study 2, appeals written by incentivized lay writers were shorter than the LLM-generated appeals (*M* = 30.0 vs. 36.0 words; *d* = 0.70). Relative to human appeals, LLM appeals were written at a higher reading level, reflected in lower Flesch Reading Ease scores (lower = harder to read; d = -1.60 and -1.56) and higher Flesch–Kincaid Grade Level scores (higher = more years of education required; d = 1.21 and 1.07, for Studies 1 and 2). Moreover, they were more lexically sophisticated, as they used longer words (mean word length d = 2.18 and 2.10), a higher proportion of long words (>6 letters; d = 1.35 and 1.48), and a more diverse vocabulary (higher type–token ratio, i.e., a higher proportion of unique words; d = 1.47 and 0.86). They were also more emotionally positive (LIWC positive tone: d = 1.22 and 1.03; overall affect d = 1.11 and 1.02), used more moral language (d = 0.86 and 1.09) and a more analytic style (d = 0.48 and 0.35), and scored lower on LIWC Authenticity (d = -0.35 and -0.40) and dictionary-word coverage (the percentage of words in a text that are contained in the LIWC dictionary; d = −1.90 and

−1.96), pointing to a more formal, curated register. Differences in negative tone were small and non-significant in both studies.

## General Discussion

People and organizations increasingly use Large Language Models (LLMs) for various text generation tasks. Emerging research indicates that LLMs have the capability to persuade humans (Bai et al., 2025; Costello et al., 2024; Schoenegger et al., 2025). Yet their potential to promote costly prosocial behavior has remained underexplored so far. We conducted two experiments testing whether LLM-generated donation appeals can lead to actual charitable giving, generating donation amounts comparable to or even higher than human-authored texts, and whether personalization to individuals' sociodemographic profiles strengthens their persuasive effectiveness.

### LLM-generated vs. Human-authored Content

In both studies, donation appeals created by LLMs generated statistically significantly higher donation amounts, higher engagement, and stronger perceptions of persuasiveness compared to those created by humans, though the absolute differences in donation amounts were small. This was the case regardless of whether human writers were academically trained in the behavioral sciences (Study 1) or U.S. lay individuals who were financially incentivized and wrote only for their own demographic group (Study 2). These results are in line with findings reported elsewhere that LLMs are comparable or even superior to human performance in persuasive communication tasks (Bai et al., 2025; Huang & Wang, 2023; Schoenegger et al., 2025). LLMs have demonstrated the ability to actively promote prosocial behavior, with effects beyond changing attitudes, but also eliciting costly actions, such as financial donations. To our knowledge, this is the first empirical study to demonstrate that LLM-generated donation appeals can elicit donations and engagement at least on par with – and in both studies statistically exceeding – those of human authors. These outcomes expand a research field dominated by concerns about LLM-related risks, including disinformation, and suggest different avenues for considering LLM technologies to increase collective well-being (Dörr et al., 2025; Goldstein et al., 2024; Spitale et al., 2023).

The reasons *why* LLM-generated donation appeals were more persuasive than human-authored ones are still speculative and require further research. Our exploratory text analyses showed that LLM-generated appeals systematically differed from human-authored appeals across a number of linguistic dimensions, including greater lexical elaborateness and more emotional and moral language. Although these analyses are descriptive and cannot establish causal mechanisms, they suggest that stylistic differences between human and LLM writing may partially contribute to the observed persuasion advantage.

Several more non-mutually exclusive explanations may account for this, too: First, LLMs may simply produce more polished, grammatically fluent text with fewer stylistic inconsistencies than human writers, a quality advantage that could translate into higher perceived credibility (Tengler & Brandhofer, 2025; Herbold et al., 2023). For example, human writers may, on average, produce text with more typos and plain language, whereas LLMs may produce output that is grammatically correct and stylistically sophisticated, which could increase the attributions of trustworthiness and professionalism to LLM-generated texts – factors shown to be important in prior persuasion research (Tengler & Brandhofer, 2025; Herbold et al., 2023; Pornpitakpan, 2004; Fogg et al., 2003).

Second, LLMs trained on vast corpora of persuasive text may hold representations of rhetorical patterns, emotional framings, and persuasion techniques that existing research has shown to be effective in eliciting prosocial behavior (Hölbling et al., 2025). For example, LLMs may have implicitly learned which combinations of urgency framing, social proof, and emotional narrative are used in prosocial contexts, based on millions of fundraising appeals, news articles, and advocacy texts. A human writer, by contrast, no matter how experienced and skilled, always draws on a comparatively narrow personal repertoire of persuasive strategies.

Third, LLMs may use morally charged and emotional semantics more effectively than humans (Carrasco-Farre, 2024), which could be the trigger for the empathic and moral motivations that drive charitable giving (Bloom, 2017; Batson, 1991). For example, LLMs may be more often able to trigger moral values such as "fairness," and, in doing so, could make the appeals feel more normatively compelling and increase engagement (Puklavec et al., 2024; Brady et al., 2017; Feinberg & Willer, 2013).

Exploring these potential mechanisms and alternative explanations represents an important task for future research. We also encourage future research to make use of other methods (e.g., field studies) to evaluate the robustness of these effects and to understand the extent to which they generalize across varied contexts and samples.

**Personalization**

Empirical evidence on the persuasive effects of personalization in LLMs is mixed (Hackenburg & Margetts, 2024a; Teeny & Matz, 2024). While some studies indicate that demographically personalized content is perceived as more persuasive than generic content (Matz, Beck, et al., 2024; Matz, Teeny, et al., 2024; Salvi et al., 2024), others report null effects (Timm et al., 2025; Hackenburg & Margetts, 2024a). In the present research, the findings regarding the role of personalization were similarly mixed: Although falsely personalized messages consistently underperformed (compared to personalized and generic messages), personalized messages resulted in higher donations than both generic and falsely personalized messages in Study 2 but not in Study 1. The better performance of personalized messages in Study 2 likely reflects greater alignment between message creators and recipients. Here, U.S.-based participants wrote messages targeting their own demographic group, whereas in Study 1, participants had to address multiple demographic groups – an obviously harder task. The findings concur with the Elaboration Likelihood Model, which predicts deeper processing for self-relevant messages (Cacioppo et al., 1986), and Social Identity Theory, which suggests that congruence with salient group identities (e.g., political orientation or religiosity) increases trust and identification with the communicator (Turner et al., 1979). In this light, our found effects of false personalization are also interpretable: When a recipient perceives a message as addressed to a mismatched recipient, the perceived relevance may drop below that of a neutral generic appeal, so the result is a net negative persuasion effect compared to generic content.

Future research is needed to explain the contextual factors and underlying psychological mechanisms by which personalization influences persuasion.

## Implications

Our results suggest that LLMs could serve as effective instruments for prosocial persuasion, specifically, in domains like fundraising. Furthermore, personalization appears to be associated with an increased likelihood of prosocial behavior even in LLM-written content. However, false personalization may be counterproductive, resulting in lower performance than no personalization at all, especially when applied to LLM-created content. Thus, practitioners and researchers are asked to create tools like LLM-driven personalized interventions, which show a good content-to-audience or audience-to-content matching (Matz, Beck, et al., 2024). For practitioners, such as fundraisers, making sure that personalization is based on accurate data should be the top priority. If the reliability of the data is in question, it is preferable to use generic content rather than risk delivering incorrectly personalized messages. Furthermore, future research needs to incorporate upcoming LLM models, integrate newly acquired empirical insights about their functioning, and include further personalization dimensions (Zettler & Strandsbjerg, 2025).

## Limitations and Future Directions

Our studies were limited to U.S. samples and to personalization based on age, gender, politics, and religiosity, and may therefore be missing broader cross-cultural variation and additional targeting factors like education (Zettler & Strandsbjerg, 2025). Both cultural background and educational attainment can influence social norms, message reflection skills, and personal capacity to engage in prosocial behavior. While our data showed an advantage of Large Language Models (LLMs) in the US context, this success may depend on the high availability of English-language and US-centric training data. In languages with lower data coverage or in cultural settings underrepresented during training, LLM outputs may lack natural fluency and normative and cognitive alignment. Such deficiencies can significantly reduce or reverse the observed persuasive advantage of LLMs over human counterparts (Rathje et al., 2024). For example, early research data points out that LLMs especially struggle to emulate the psychological profiles and moral intuitions of individuals outside of the Western, Educated, Industrialized, Rich, and Democratic (WEIRD) societies (Atari et al., 2023). Future research could consider a different set of demographic, cultural, and psychosocial factors.

Furthermore, the prosocial behavior was measured in the present studies via the allocation of a bonus provided by the researchers, rather than participants' own money, which may have reduced the weight of the donation decision and could limit the generalizability of the findings in this regard.

A further limitation concerns the single-page presentation of all six posts simultaneously: While this was intended to minimize carryover effects and standardize viewing conditions, it could have invited participants to compare posts directly against each other rather than evaluating each in isolation. This reduces ecological validity relative to social media scrolling, where charity appeals would typically be encountered sequentially, interspersed with other, unrelated content, and less frequently than in the present studies.

The present studies provide a first benchmark with two human baselines: academically trained writers and incentivized lay writers from the target population. A natural next step could be to additionally compare LLM-generated appeals with professional, niche-specialized fundraising copywriters and niche-specialized nonprofit communication services, especially if charities decide that such expertise can be afforded by them and/or complemented, scaled, or substituted by LLMs.

Also, in the present studies, communication with the charities was unidirectional, with recipients unable to respond - unlike interactive contexts such as online support chats. Future work could extend this work by examining trait-based personalization in interactive LLM-mediated dialogues (Costello et al., 2024). In this context, recent work suggests that LLMs are not only able to respond to counterarguments but also able to actively produce consensus via bridging different viewpoints into broader, more widely acceptable statements (e.g., the "Habermas Machine"; Tessler et al., 2024). However, the one-sided message format used here will likely remain the predominant paradigm for mass-market persuasive communication (e.g., via social media posts and paid digital ads) for the next few years.

A further study direction could be to analyze the linguistic characteristics of LLM-generated and human-authored appeals more mechanically: As a first step in this direction, we exploratively analyzed the appeals on a set of text characteristics (length, readability, lexical sophistication, and emotional/moral content). In the future, for example, Natural Language Processing (NLP) or human coding approaches could clarify whether the observed content-source advantage reflects substantive rhetorical differences, greater fluency, or more subtle textual properties.

The message authorship in our studies was undisclosed, and investigating whether the effectiveness of LLM-generated appeals persists when recipients are aware of "AI authorship" represents an important route for future studies, as such disclosure obligations are increasingly being discussed politically and have been recently shown to reverse the LLM advantage (Laux et al., 2024; Yin et al., 2024). Specifically, studies show that disclosing LLM authorship often lowers the perceived value of LLM-generated content, compared to blind evaluations. Even in domains such as personal advice and acute emotional support, LLM-generated responses receive high ratings when their source is unknown, but ratings diminish or reverse once recipients learn the content was generated by an LLM (Osborne & Bailey, 2025; Rubin et al., 2025; Yin et al., 2024).

**Conclusion**

Our data show that LLM-generated appeals can perform better than human-written appeals in promoting prosocial donation allocations in a controlled experimental setting. At the same time, our findings caution that personalization is only useful when implemented accurately, as mismatched personalization might backfire and undermine the very effectiveness it was meant to foster. As LLMs become more developed and widely deployed, understanding if and how they can influence prosocial behavior will remain important for future studies.

**Acknowledgments**

Besides all participants of both studies and human writers on Prolific of Study 2, we thank for their contribution in writing the human-authored posts for Study 1: Mariam Bolkvadze, Qian Chen, Jennifer Chen, Hieu Doan, Jonas Festor, Riccardo Loconte, Tom Nyhoff, Lucca Pfründer, Lars Probst, Sanne Peereboom, Ivo Snels, Ngoc Anh Tran, Stefana Vida, Weng Ao, and Jari Zegers.

# Appendix

Appendix 1

*Post Content Wrapper*

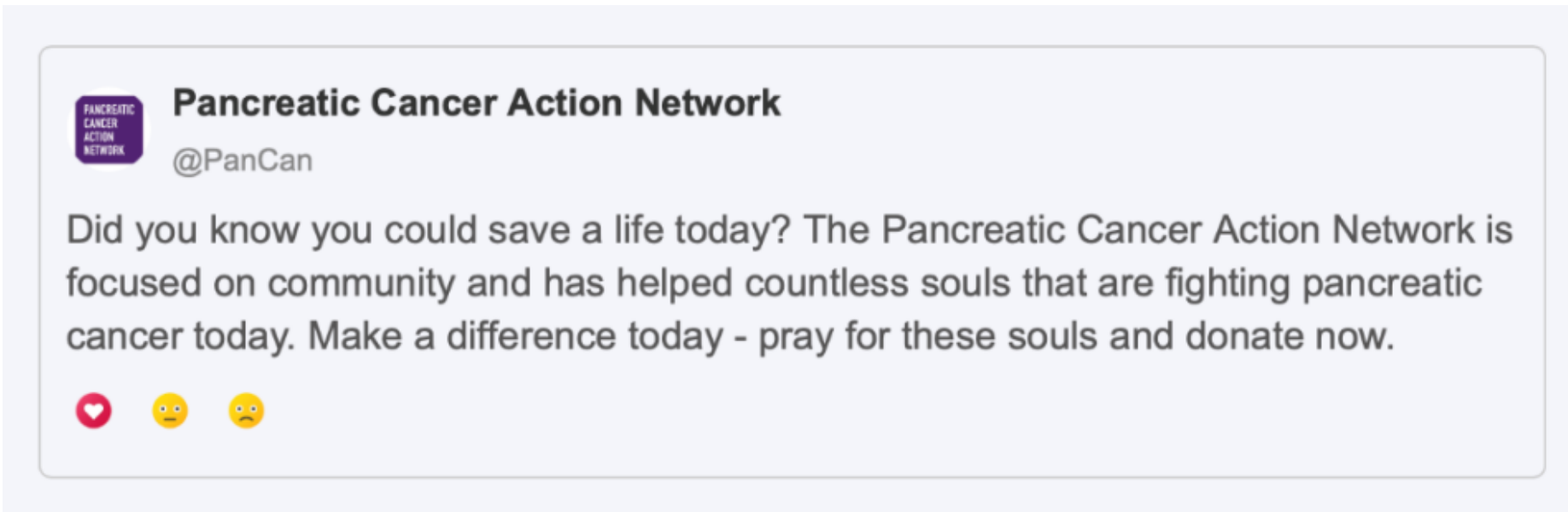

Appendix 2

*Donation Amounts (in $-cent steps) in Study 1*

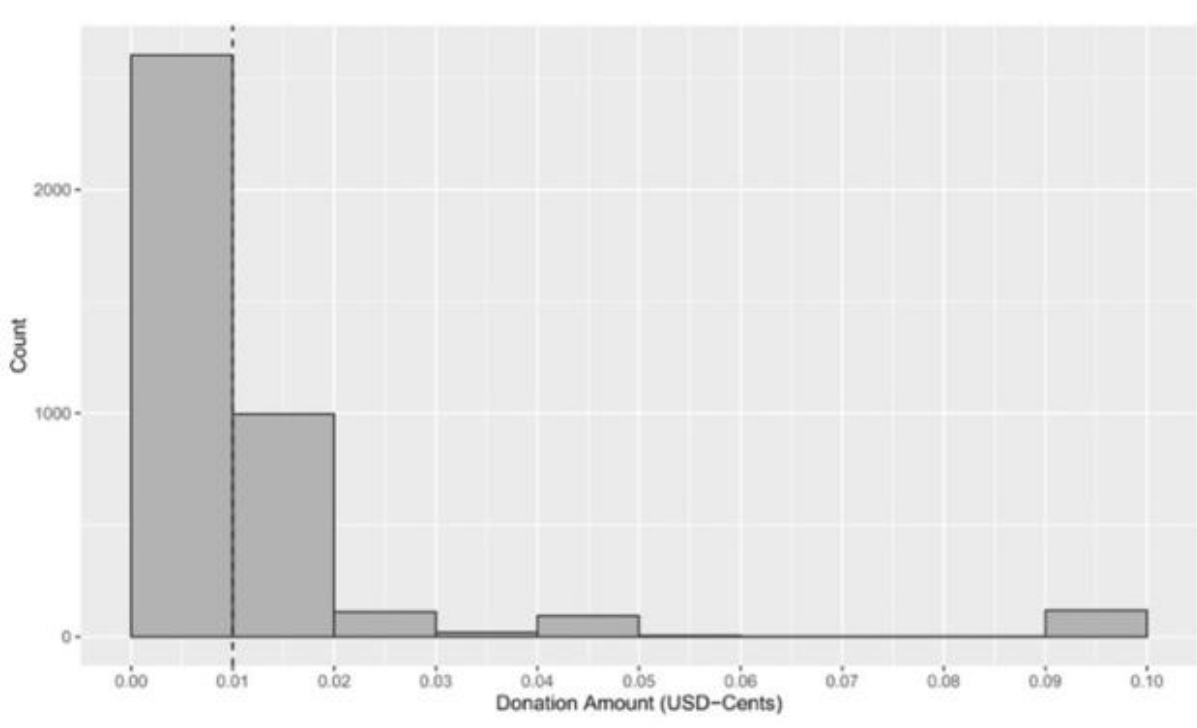

Appendix 3

Average donation per person in Study 1

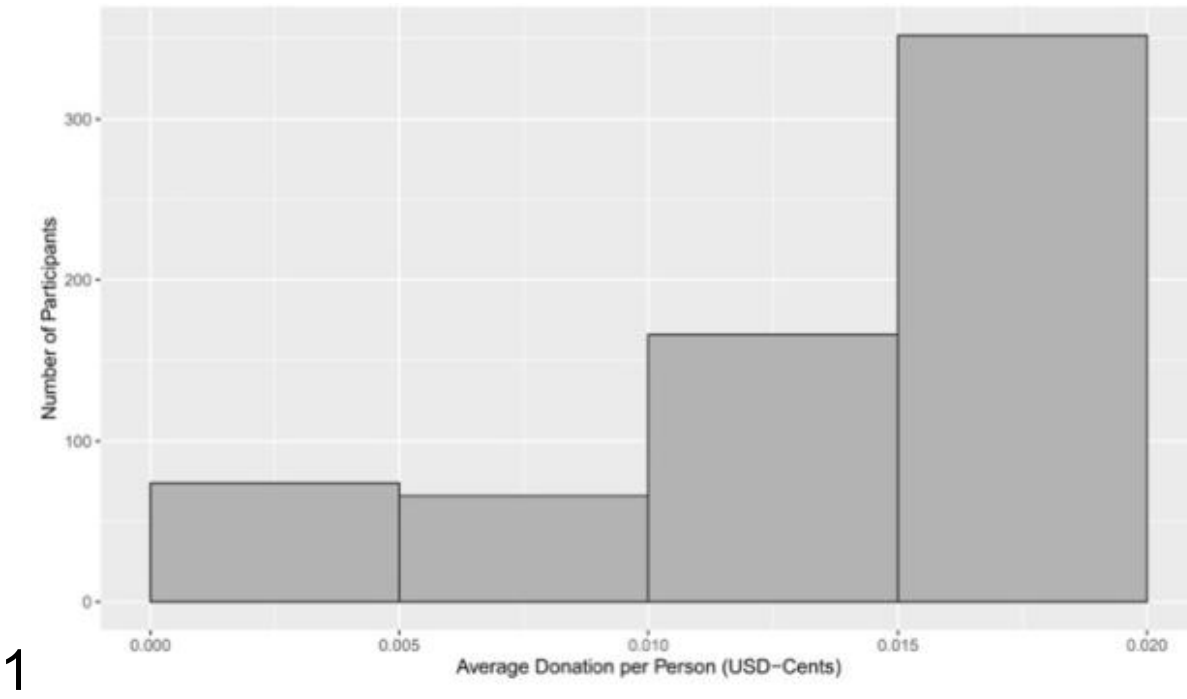

Appendix 4

*Participants per socio-demographic cluster in Study 1*

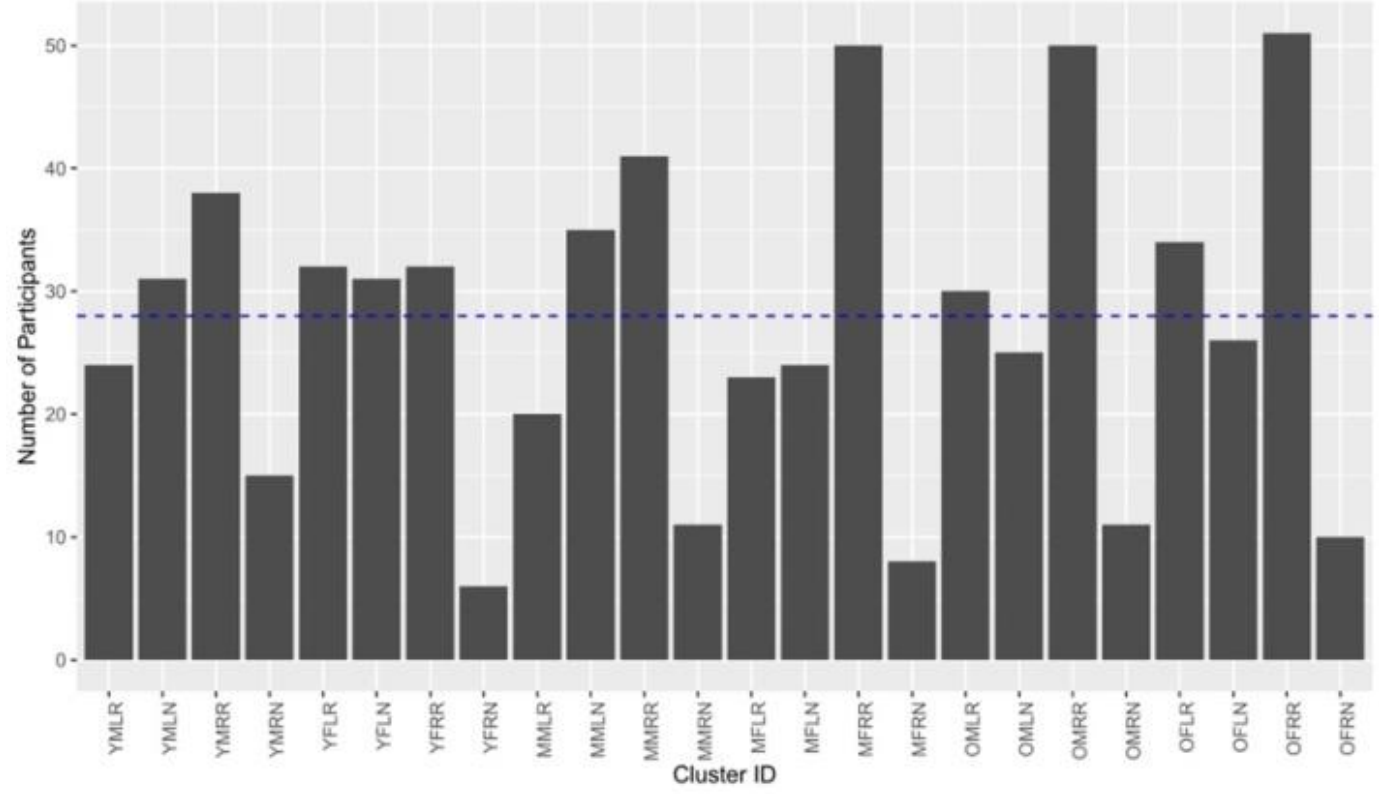

Appendix 5

*Distribution of Post Engagement (Dislike/Neutral/Like) in Study 1*

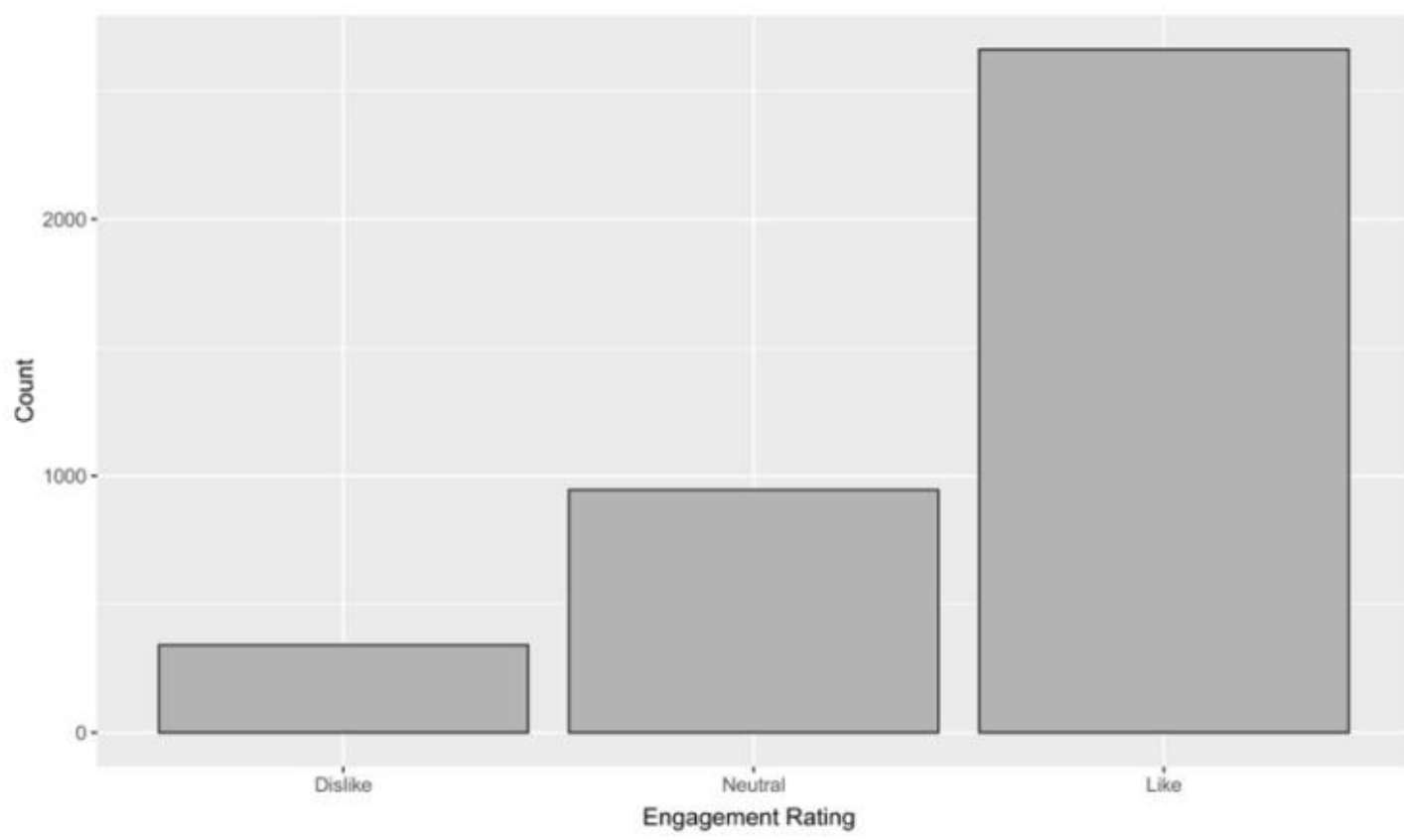

Appendix 6

*Donation Distribution (Study 1)*

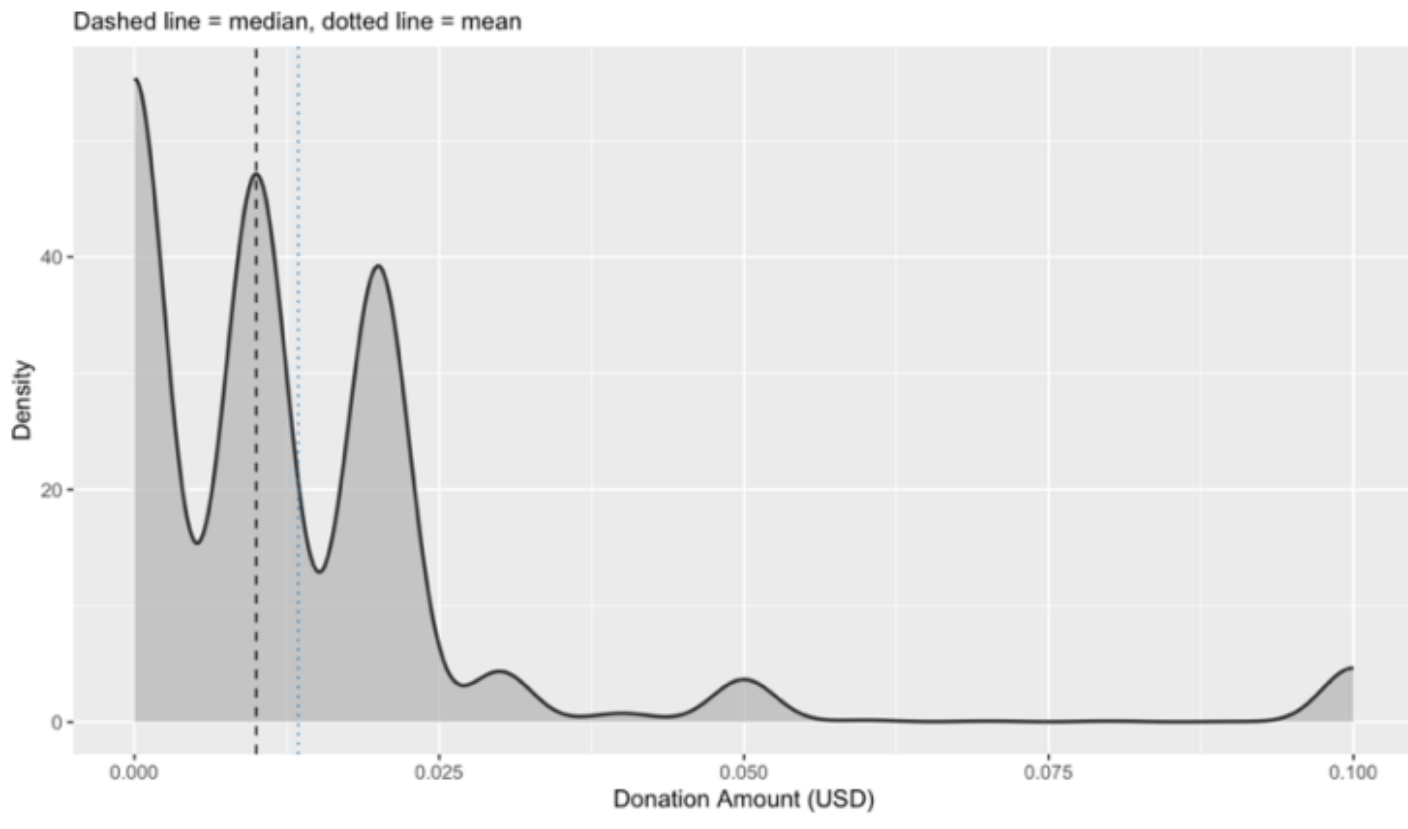

## Appendix 7

*Tukey-adjusted pairwise/post-hoc comparisons (Study 1 & 2)*

| | Source | Contrast | *SE* | *p* |
|---|---|---|---|---|
| Study 1 — Donation Amount | | | | |
| | Human | generic – personalized | .0010 | .887 |
| | Human | generic – falsely personalized | .0010 | .241 |
| | Human | personalized – falsely personalized | .0010 | .095 |
| | LLM | generic – personalized | .0010 | .765 |
| | LLM | generic – falsely personalized | .0010 | .446 |
| | LLM | personalized – falsely personalized | .0010 | .136 |
| Study 1 — Post Engagement | | | | |
| | Human | generic – personalized | .027 | .039 |
| | Human | generic – falsely personalized | .027 | .990 |
| | Human | personalized – falsely personalized | .027 | .055 |
| | LLM | generic – personalized | .027 | .999 |
| | LLM | generic – falsely personalized | .027 | < .001 |
| | LLM | personalized – falsely personalized | .027 | < .001 |
| Study 1 — Persuasiveness | | | | |
| | Human | generic – personalized | .059 | .905 |
| | Human | generic – falsely personalized | .059 | .622 |
| | Human | personalized – falsely personalized | .059 | .365 |
| | LLM | generic – personalized | .059 | .170 |
| | LLM | generic – falsely personalized | .059 | < .001 |
| | LLM | personalized – falsely personalized | .059 | < .001 |
| Study 2 — Donation Amount | | | | |
| | Human | generic – personalized | .0012 | .151 |
| | Human | generic – falsely personalized | .0012 | .942 |
| | Human | personalized – falsely personalized | .0012 | .272 |
| | LLM | generic – personalized | .0012 | .059 |
| | LLM | generic – falsely personalized | .0012 | .893 |
| | LLM | personalized – falsely personalized | .0012 | .017 |
| Study 2 — Post Engagement | | | | |
| | Human | generic – personalized | .029 | .008 |
| | Human | generic – falsely personalized | .029 | .051 |
| | Human | personalized – falsely personalized | .029 | .776 |
| | LLM | generic – personalized | .029 | .685 |
| | LLM | generic – falsely personalized | .029 | < .001 |
| | LLM | personalized – falsely personalized | .029 | < .001 |
| Study 2 — Persuasiveness | | | | |
| | Human | generic – personalized | .067 | .142 |
| | Human | generic – falsely personalized | .066 | .824 |
| | Human | personalized – falsely personalized | .066 | .392 |
| | LLM | generic – personalized | .066 | .992 |
| | LLM | generic – falsely personalized | .066 | < .001 |
| | LLM | personalized – falsely personalized | .066 | < .001 |

Appendix 8

*Donation Amounts (in $-cent steps) in Study 2*

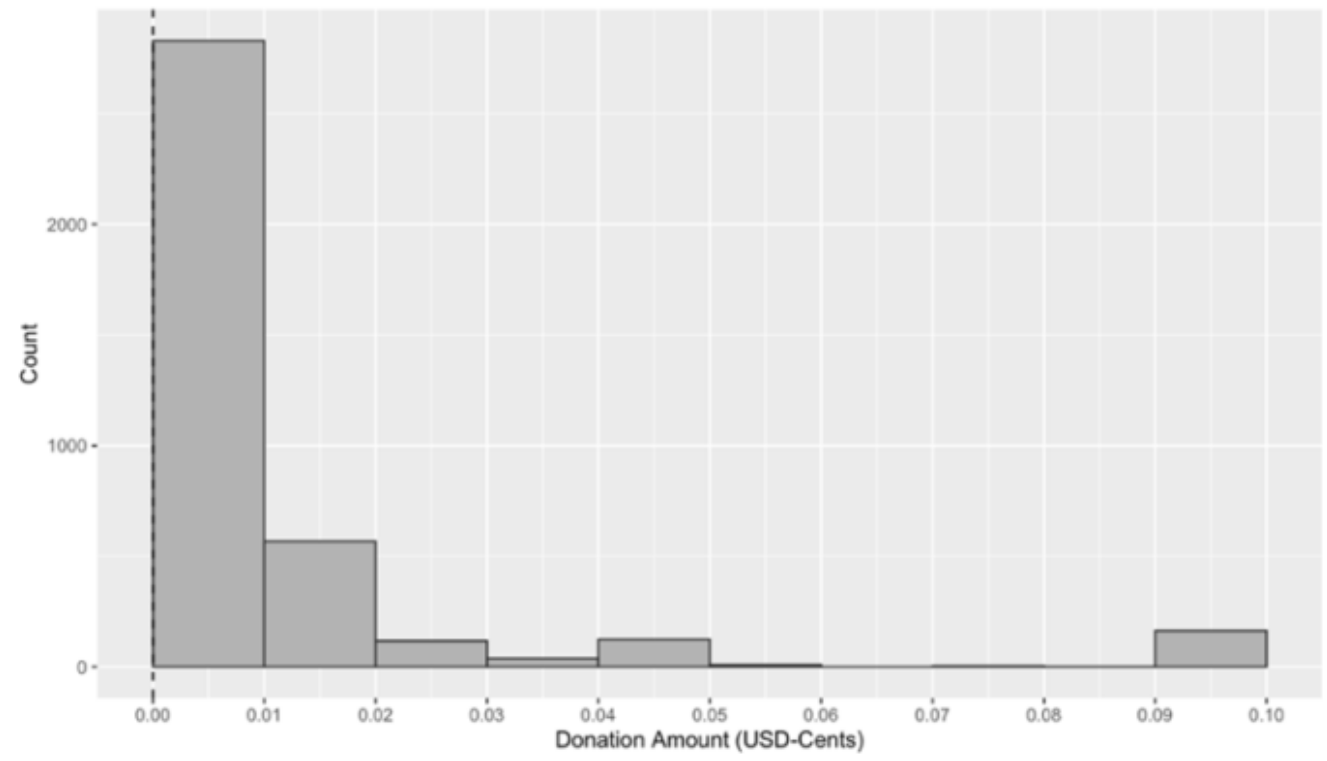

Appendix 9

*Average donation per person of Study 2*

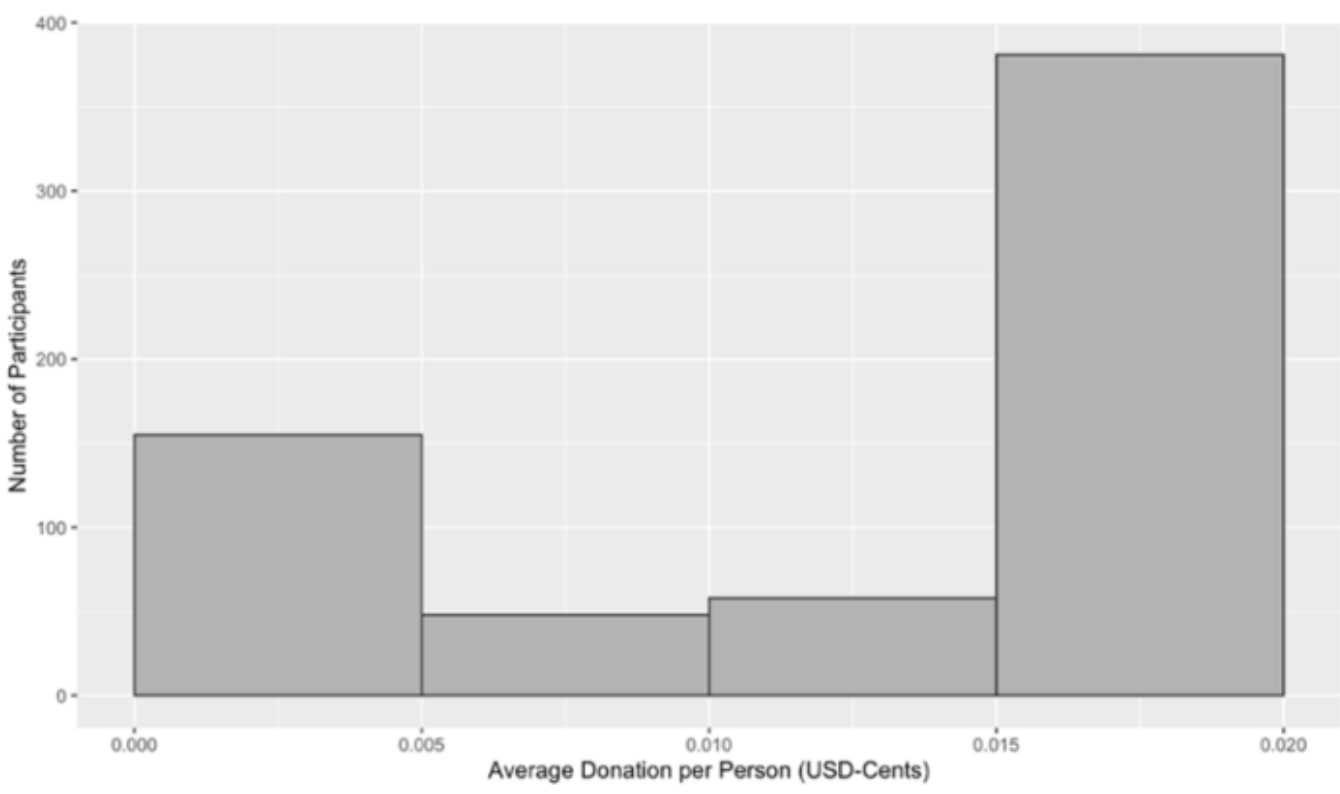

Appendix 10

*Participants per cluster of Study 2*

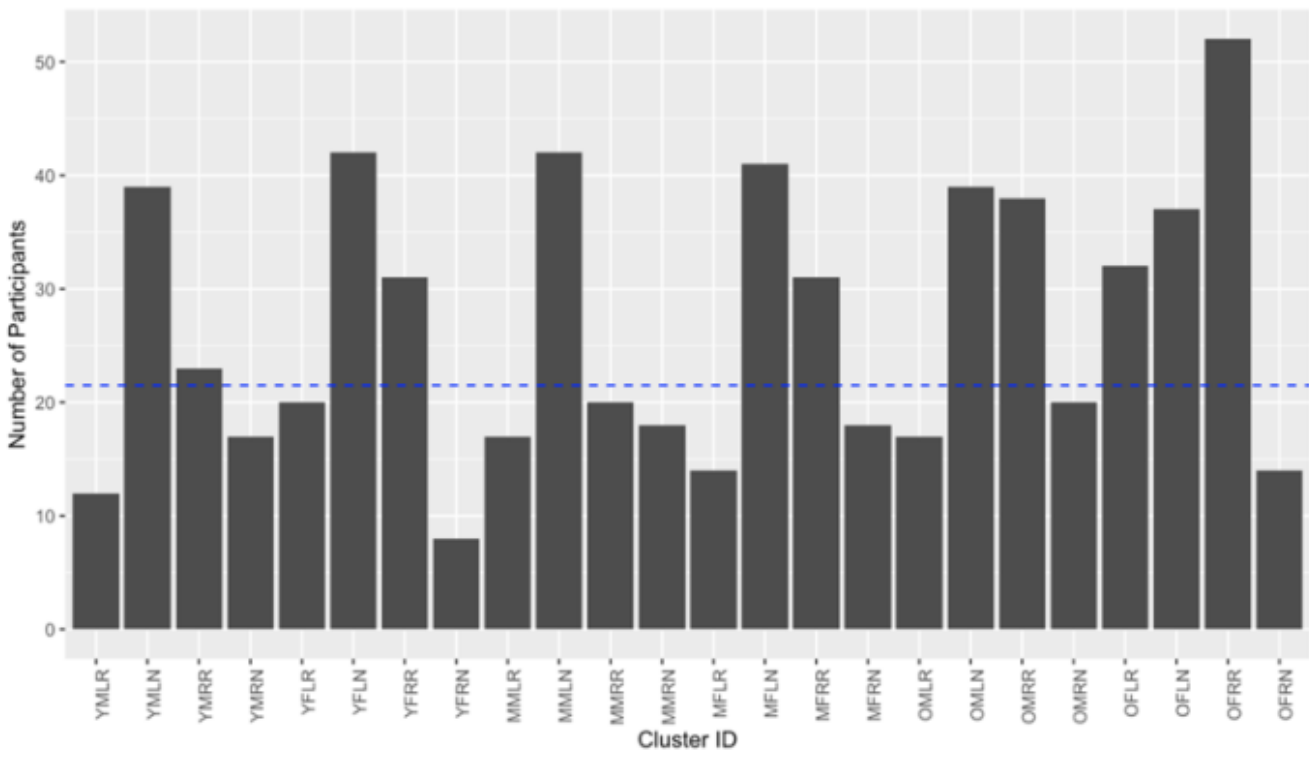

Appendix 11

*Distribution of Post Engagement (Dislike/Neutral/Like) of Study 2*

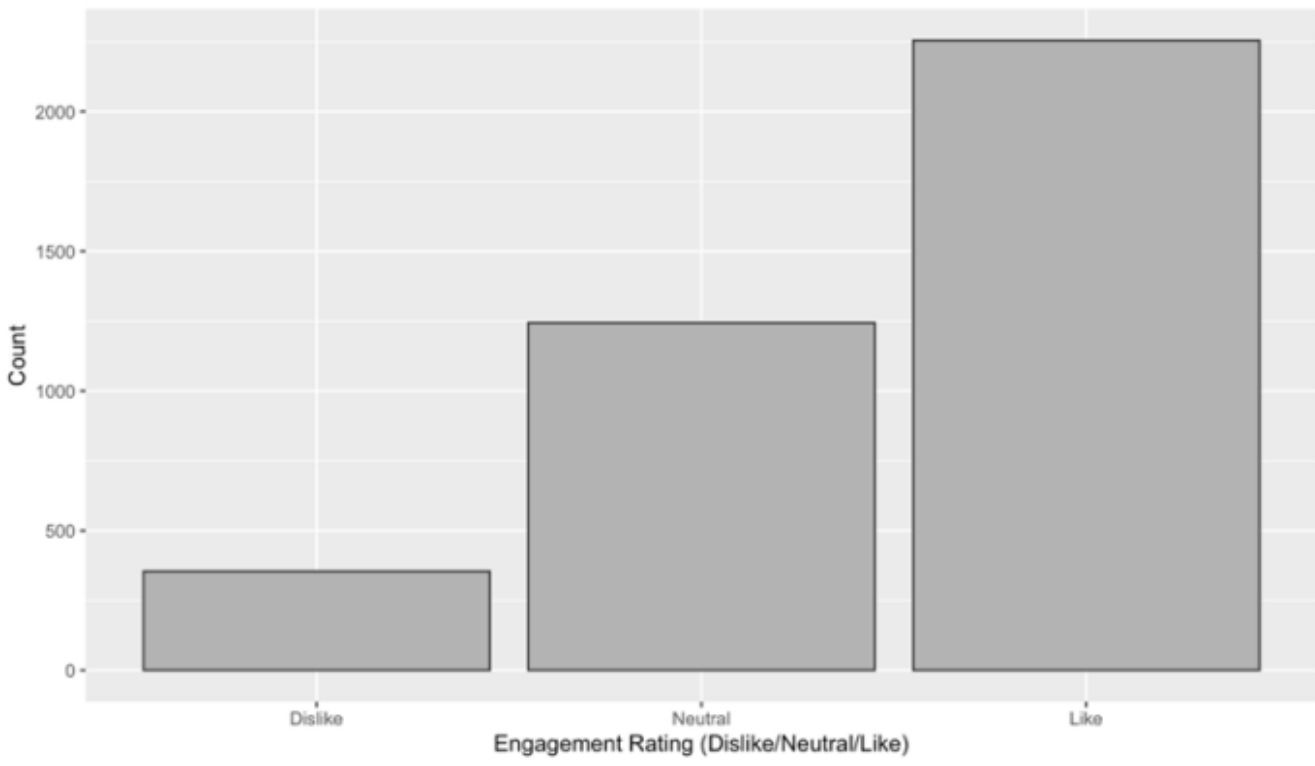

Appendix 12

*Donation Distribution (Study 2)*

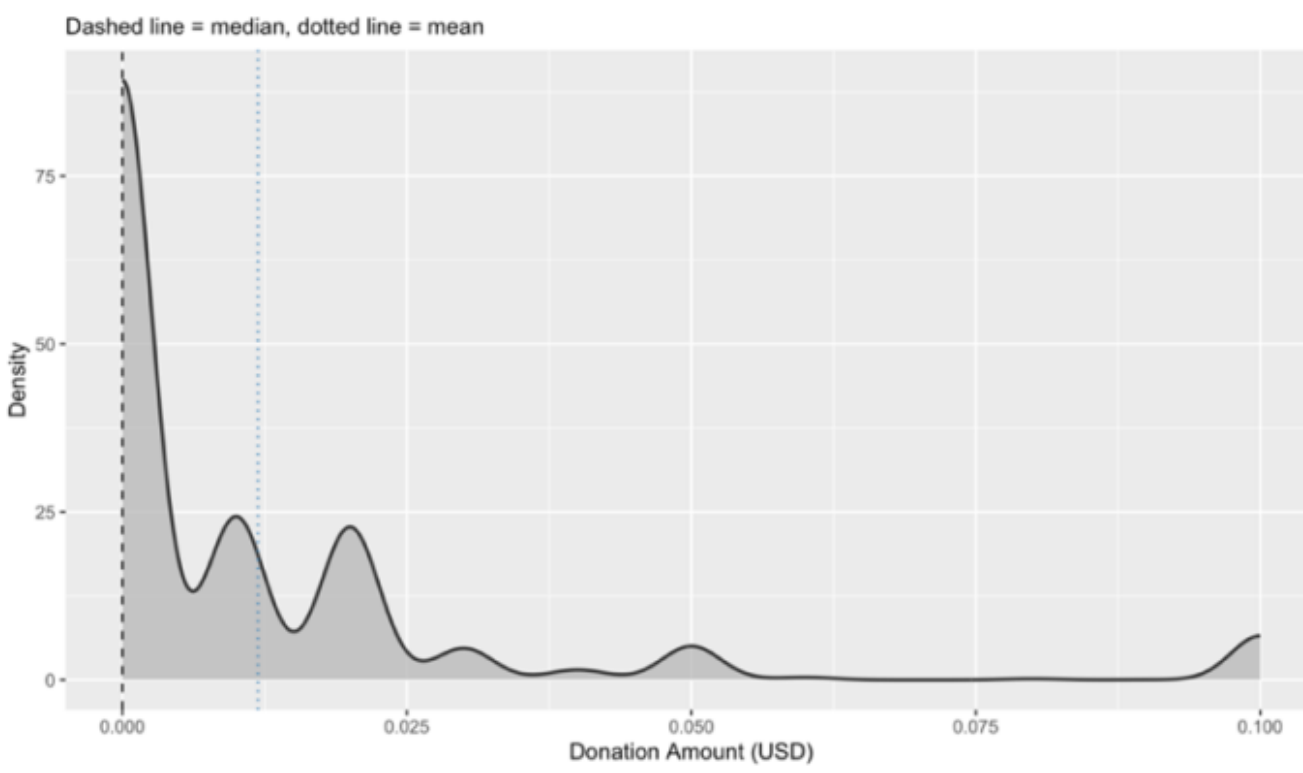